\documentclass{aa}  

\usepackage[varg]{txfonts}
\usepackage{graphicx}
\usepackage{booktabs}
\usepackage{array}
\usepackage{hyperref}

\usepackage{float}

\begin{document} 

\title{Angular momentum evolution in the CIELO simulations}
\subtitle{I. Temporal evolution of gas--stellar misalignments and baryonic merger timing}

\author{Catalina Casanueva-Villarreal\inst{1,2} \and Patricia B. Tissera\inst{1,2} \and Nelson Padilla\inst{3,4} \and Lucas Bignone\inst{5} \and Jenny González-Jara\inst{1,2} \and Brian Tapia-Contreras\inst{1,2} \and Susana Pedrosa\inst{5} }

\institute{Instituto de Astrofísica, Pontificia Universidad Católica de Chile, Av. Vicuña Mackenna 4860, Santiago, Chile \and Centro de Astro-Ingeniería, Pontificia Universidad Católica de Chile, Av. Vicuña Mackenna 4860, Santiago, Chile
        \and
            Instituto de Astronomía Teórica y Experimental (IATE), CONICET-Universidad Nacional de Córdoba, Laprida 854, X5000BGR, Córdoba, Argentina
        \and 
            Observatorio Astronómico de la Universidad Nacional de Córdoba, Laprida 854, X5000BGR, Córdoba, Argentina
        \and Instituto de Astronomía y Física del Espacio, CONICET-UBA, 1428 Buenos Aires, Argentina}

\date{Received Month Day, Year; accepted Month Day, Year}

\abstract
{Gas--stellar kinematic misalignments trace how galaxies assemble and reorient their angular momentum. Although well documented in the Local Universe, their continuous evolution over the lifetime of a galaxy remains largely unexplored.}
{We aim to characterise the pathways linking aligned, misaligned, and counter-rotating phases, assessing how gas accretion channels and mergers drive reorientation.}
{We analysed 44 central galaxies from the \emph{Chemo-dynamIcal propertiEs of gaLaxies and the cOsmic web} (CIELO) simulations from $z=3.5$ to $z=0$. We defined kinematic episodes using the intrinsic angle, $\psi$, between the star-forming (SF) gas and stellar angular momentum vectors. We interpreted these histories by tracking accreted gas origins (smooth accretion, stripped material, mergers) and evaluating paired statistical contrasts.}
{Nearly 80\% of the simulated galaxies are aligned at $z=0$, yet 91\% experience at least one non-aligned episode. Although 19\% of non-aligned episodes last $>2$~Gyr, within-galaxy comparisons show their durations do not differ significantly from aligned episodes. Abrupt changes in $\psi$ coincide with intervals where accreted gas dominates the pre-existing SF reservoir and is highly tilted relative to pre-existing stars. During transitions, the median mass ratio of accreted to pre-existing SF gas rises from 0.57 to 2.14, and the median angular offset increases from $21.2^\circ$ to $64.6^\circ$. Minor and major mergers accompany aligned$\rightarrow$non-aligned and non-aligned$\rightarrow$aligned transitions across mass ratios and orbits. Those near abrupt transitions generally enter hosts whose stars and SF gas are already more misaligned at infall.}
{In the simulated sample, abrupt changes in $\psi$ coincide with newly added gas becoming dynamically important relative to the pre-existing SF reservoir and highly tilted relative to the pre-existing stars. Minor and major mergers occur near these transitions, while subsequent evolution reflects the balance between replacement and retention of the central SF gas reservoir.}


\maketitle

\section{Introduction}

Understanding how galaxies acquire and redistribute angular momentum is central to galaxy formation models. In the standard Lambda cold dark matter ($\Lambda$CDM) framework, gas cools inside dark-matter haloes while retaining a substantial fraction of its specific angular momentum, giving rise to a rotationally supported cold gas structure. Therefore, stars formed \emph{in situ} inherit the kinematics of their parent gas \citep{Fall_1980,Mo_1998}. However, gas and stars respond differently to accretion, torques, feedback, and mergers, enabling their angular momentum vectors to decouple \citep[e.g.,][]{Pedrosa-Tissera_2015,Teklu_2015}. The existence of gas--stellar kinematic misalignments therefore provides a direct probe of how galaxies accrete, process, and reorient their gas and stellar components.

Early evidence for this decoupling came from observations of individual galaxies hosting polar gas rings or counter-rotating components \citep[e.g.,][]{Ulrich_1975,Bertola_1992,Rubin_1992,Kannappan_2001}. Integral-field spectroscopy (IFS) later showed that such configurations are common and depend strongly on galaxy type. In ATLAS$^{3\mathrm{D}}$, \citet{Davis_2011} found that $36 \pm 5\%$ of fast-rotating early types host kinematically misaligned ionised gas. In SAMI, \citet{Bryant_2019} found an overall misalignment fraction ($>30^{\circ}$) of $\sim 11\%$, which rises to $45 \pm 6\%$ in early types (spheroids embedding gas discs) and falls to $5 \pm 1\%$ in spirals. Using $\sim 4500$ galaxies from MaNGA, \citet{Duckworth_2020} corroborated this strong morphological dependence. More recent work has linked misaligned gas acquisition to lenticular (S0) growth \citep{Zhou_2024}, central rejuvenation \citep{Zhou_2022}, mergers \citep{Li_2021,Ristea_2022}, and externally supplied gas reservoirs shaped by the large-scale structure \citep{Bao_2025}.

Theoretical and numerical work consistently favours a multi-channel origin for gas--stellar kinematic misalignments. Semi-analytic modelling linked misaligned gas in early-type galaxies to late external accretion through cooling from misaligned hot haloes, filamentary inflow, and minor mergers \citep{Lagos_2015}. In a single simulated early-type galaxy, \citet{vandeVoort_2015} showed how externally acquired gas can build a long-lived misaligned disc. Population-level analyses in large cosmological simulations have shown that multiple channels can contribute. In Horizon-AGN, \citet{Khim_2020a} found that misalignments are associated with mergers, interactions, environment, and secular evolution, while in EAGLE, \citet{Casanueva_2022} showed that environment can trigger misalignment while internal structure regulates how efficiently galaxies realign. Higher-resolution studies and analyses of restricted sub-populations have further clarified some of the conditions under which strong decoupling can persist. In FIREbox, \citet{Cenci_2023} argued that a merger-triggered starburst can deplete the original co-rotating gas and allow newly accreted retrograde gas to dominate the kinematics. In NewHorizon, \citet{Han_2024} and \citet{Peirani_2025} highlighted gas--gas angular momentum cancellation, staged coexistence of pre-existing and newly acquired gas, and progressive reservoir replacement as key ingredients of strong misalignment and counter-rotation. However, at the population level, \citet{Baker_2025a} showed in EAGLE that many unstable misalignments are short-lived, with median relaxation times of $\sim0.5\mathrm{\,Gyr}$, although continued inflow, low star-forming (SF) gas fractions, and dense central environments can prolong them to $\gtrsim1\mathrm{\,Gyr}$. 

Taken together, these studies suggest that the outcome depends not only on the external perturbation but also on the state of the pre-existing gaseous reservoir. At the same time, they probe different aspects of the problem, from individual case studies and restricted sub-populations to sample-level incidence and lifetime statistics.

What remains missing is a comprehensive statistical framework that reconstructs repeated transitions between aligned, misaligned, and counter-rotating states along the continuous evolutionary path of galaxies. While observations tightly constrain the demographics of decoupled systems in the Local Universe, their single-epoch nature cannot recover the potential continuous paths linking these classes. Existing hydrodynamical simulations have instead tended to emphasise either detailed case studies of one or a few systems, restricted sub-populations, or sample-level incidence rates and relaxation times. A uniform, population-level analysis that reconstructs those long-term paths within a single selected sample and connects them directly to gas supply, gas origin, and merger timing across multiple epochs is therefore still lacking. Resolving the time evolution is essential to distinguish brief perturbations from stable, long-lived reorientations and to reconstruct the histories hidden behind present-day kinematic classes.

Addressing this problem requires simulations that resolve both the internal structure of galaxies and the external accretion that perturbs them. Cosmological zoom-in simulations are well suited to this task because they combine resolved baryonic structures with explicit tracking of individual accretion and merger events. In this first study of gas--stellar misalignments in the \emph{Chemo-dynamIcal propertiEs of gaLaxies and the cOsmic web} 
CIELO zoom-in simulations \citep{Tissera_2025}, we follow the time evolution of the fiducial 3D angle, $\psi$, between the angular momentum vectors of the SF gas and the stellar component in 44 central galaxies selected at $z=0$. We use the snapshot sequence from $z=3.5$ to $z=0$ to define aligned and non-aligned episodes, where non-aligned denotes the combined set of misaligned and counter-rotating states. After analysing six representative systems, we extend this analysis to the full sample. This work adds to the literature not only by identifying misaligned gas and its association with mergers and accretion, but also through a novel reconstruction of alignment-to-misalignment transitions within the same galaxies, combined with the tracking of individual gas particles as they enter the central SF reservoir.

This paper is organised as follows. Section~\ref{sec:sim_kin} describes the simulations, sample selection, angle definitions, and episode construction. In Sect.~\ref{sec:case_pathways}, we present six representative systems and the five representative pathways they illustrate into and out of misalignment. Section~\ref{sec:global_stats} extends this framework to the full 44-galaxy sample using paired statistics, gas-supply and origin diagnostics, and merger-context tests. Finally, we discuss the implications of our results in Sect.~\ref{sec:discussion} and conclude in Sect.~\ref{sec:conclusions}.

\section{Simulations, sample, and angle definitions}
\label{sec:sim_kin}

\subsection{The CIELO suite}
\label{subsec:cielo}

Our analysis is based on the CIELO suite of chemo-hydrodynamical cosmological zoom-in simulations \citep{Tissera_2025}. The simulations assume a $\Lambda$CDM cosmology with $\Omega_{\rm m}=0.317$, $\Omega_{\Lambda}=0.683$, $\Omega_{\rm b}=0.049$, $h=0.6711$, $\sigma_{8}=0.834$, and $n_{\rm s}=0.962$ \citep{PlanckCollaborationXVI_2014}. The initial conditions were generated with \textsc{Music} \citep{Hahn-Abel_2011} from two dark-matter-only parent simulations. The first is a 100~$h^{-1}$~Mpc box used to select Local Group (LG) analogues as paired environments according to the relative positions and velocities of their two primary haloes. The second is a periodic box of side length 50~$h^{-1}$~Mpc used to define the Pehuen sample, whose target haloes span groups, filaments, walls, and voids. Haloes in the parent simulations were identified with \textsc{Rockstar} \citep{Behroozi_2012}.

The zoomed regions were resimulated with baryons using a modified version of \textsc{Gadget-3} that includes a multiphase interstellar medium, metal-dependent radiative cooling, stochastic star formation, supernova feedback, and chemical enrichment \citep{Scannapieco_2005,Scannapieco_2006}. We use both resolution levels available in CIELO. The L12 runs have dark-matter particle masses of $m_{\rm dm}=1.36\times10^{5}\,\mathrm{M}_\odot\,h^{-1}$ and initial gas particle masses of $m_{\rm gas}=2.1\times10^{4}\,\mathrm{M}_\odot\,h^{-1}$, while the L11 runs have $m_{\rm dm}=1.28\times10^{6}\,\mathrm{M}_\odot\,h^{-1}$ and $m_{\rm gas}=2.0\times10^{5}\,\mathrm{M}_\odot\,h^{-1}$. The gravitational softenings of gas and stellar particles are 250~pc in L12 and 400~pc in L11. The corresponding dark matter softenings are 500 and 800~pc. All runs were evolved from $z\sim60$ to $z=0$ and stored in 128 snapshots at a typical cadence of $\sim0.17$~Gyr. The chemo-dynamical subgrid physics in this suite has been shown to successfully reproduce the diverse metallicity profiles of SF gas \citep{TapiaContreras_2025} as well as the assembly channels and chemical fingerprints of stellar haloes \citep{GonzalezJara_2025_chemhalos} and bulges \citep{MunozEscobar_2026}.

\subsection{Sample selection}
\label{subsec:sample}

We select central galaxies at $z=0$ from the CIELO catalogue that are numerically well resolved for robust angle measurements. We adopt mass thresholds of $M_\star > 10^8\mathrm{\,M_\odot}$ for the L12 runs and $M_\star > 10^9\mathrm{\,M_\odot}$ for the L11 runs, ensuring at least $\sim 3\times10^{3}$ stellar particles. Following the observational definition of the optical radius \citep{Persic_1996}, we define its stellar-mass analogue, $r_{\rm opt,\star}$, as the radius enclosing 83\% of the stellar mass. We measure all global quantities within $2 r_{\rm opt,\star}$, an aperture that robustly encompasses the main stellar body while reducing the kinematic influence of the outskirts. We additionally require $\geq 100$ SF gas particles within $2 r_{\rm opt,\star}$ and a stellar-gas centre-of-mass separation $< 2$~kpc to ensure the gas reservoir is dynamically coupled to the central galaxy.

The final sample comprises 14 L11 galaxies and 30 L12 galaxies (see full breakdown in Appendix Table~\ref{tab:final_sample_detail}). The galaxies span $\log(M_{\star}/\mathrm{M}_\odot)=8.0$--$10.7$ (median $9.5$). The retained galaxies lie well above the minimum SF gas threshold, containing from a few hundred to $>4\times10^{4}$ such particles within $2 r_{\rm opt,\star}$ at $z=0$ (median $\sim 4.6\times10^{3}$). The resolution and region labels are used descriptively, as no dedicated convergence or environment-controlled analysis is attempted here.

\subsection{Component definitions and angular-offset estimators}
\label{subsec:misalignment_metric}

We define SF gas as the cold ($T<1.5\times10^{4}$~K) and dense ($\rho>7\times10^{-26}$~g~cm$^{-3}$) phase triggering the CIELO multiphase star-formation model \citep{Tissera_2025}. At each snapshot, a common centre is obtained by applying a shrinking-sphere algorithm to all stellar particles assigned to the central subhalo, starting from the \textsc{subfind} centre \citep{Power_2003}. Gas and stellar kinematics are measured within $2\,r_{\rm opt,\star}$ around this centre. For any component, we compute the total angular momentum vector as
\begin{equation}
\mathbf{J}=\sum_i m_i\,(\mathbf{r}_i\times\mathbf{v}_i),
\label{eq:angular_momentum}
\end{equation}
where $m_i$ is the mass, and $\mathbf{r}_i$ and $\mathbf{v}_i$ are the position and velocity vectors of the $i$-th particle. Its corresponding unit vector is $\hat{\mathbf{J}}=\mathbf{J}/|\mathbf{J}|$.

Our fiducial intrinsic offset is the 3D angle between the SF gas and all stars:
\begin{equation}
\psi \equiv \cos^{-1}\!\left(\hat{\mathbf{J}}_{\rm SF,\,\leq 2r_{\rm opt,\star}}\cdot\hat{\mathbf{J}}_{\star,\,\leq 2r_{\rm opt,\star}}\right).
\end{equation}

Both angular momentum vectors are computed from all SF gas and stellar particles within the spherical aperture of radius $2r_{\rm opt,\star}$; no morphological selection or disc plane cut is applied.

We classify galaxies as ``aligned'' for $\psi<30^\circ$, ``misaligned'' for $30^\circ\leq\psi<150^\circ$, and ``counter-rotating'' for $\psi\geq150^\circ$. This $30^\circ$ boundary follows conventions widely used in observational and simulation studies of gas--stellar misalignment, where it is typically applied to projected kinematic position-angle offsets \citep{Davis_2011,Bryant_2019,Casanueva_2022}. In our intrinsic 3D analysis, this threshold serves purely to identify physically misaligned systems rather than to account for observational errors. At $z=0$, this classification yields 35 aligned, 6 misaligned, and 3 counter-rotating galaxies. These classes overlap substantially in stellar mass: aligned, misaligned, and counter-rotating galaxies span $\log(M_{\star}/\mathrm{M}_\odot) \sim 8.0$--$10.7$, $8.2$--$10.6$, and $9.3$--$10.2$, respectively (Appendix Table~\ref{tab:final_sample_detail}). Although the three counter-rotating systems tend to be more massive, this small sample precludes formal mass dependence claims.

To evaluate the stability of these fiducial classes, we compare $\psi$ against two alternative angle definitions, applying the same $30^\circ$ and $150^\circ$ thresholds for consistency. First, to test whether our classification depends on the selected stellar population, we compute the intrinsic 3D angle replacing all stars with a kinematically selected subset of disc stars ($\psi_{\rm 3D,disc}$). These disc stars are identified from binding energy, $E$, and circularity, $\epsilon=j_z/j_{z,\max}(E)$, where $j_z$ is the specific angular momentum parallel to the galaxy rotation axis \citep[e.g.][]{Tissera_2012,Pedrosa-Tissera_2015}. Replacing all stars with disc stars proves highly stable: it changes the class of only one borderline system, P3--gx298 ($\psi=149.6^\circ$), which crosses the $150^\circ$ threshold to become counter-rotating. A second near-boundary case, P7--gx200 ($\psi=29.4^\circ$), remains securely aligned ($\psi_{\rm 3D,disc}=29.0^\circ$).

Second, to assess sensitivity to viewing geometry and dimensionality, we calculate the 2D angles between the angular momentum vectors projected onto the three orthogonal planes of the simulation box ($xy$, $xz$, and $yz$). As galaxies are randomly oriented with respect to the cosmological grid, these projections provide three orthogonal viewing directions per system. In observational IFS studies, the relative gas--stellar orientation is typically expressed as the difference between the projected stellar and gaseous kinematic position angles, $\Delta \mathrm{PA}$, measured from velocity fields \citep[e.g.][]{Davis_2011,Barrera-Ballesteros_2014,Bryant_2019,Duckworth_2020,Zinchenko_2023}. Since kinematic position angles trace the projected angular momentum \citep{Krajnovic_2006,Davis_2011}, we use our 2D projections as direct proxies for the observational $\Delta \mathrm{PA}$. For a projected plane $p\in\{xy,xz,yz\}$, we define:
\begin{equation}
\Delta \mathrm{PA}_{\rm 2D}^{(p)} \equiv \cos^{-1}\!\left(\hat{\mathbf{J}}^{\,(p)}_{\rm SF,\,\leq 2r_{\rm opt,\star}}\cdot\hat{\mathbf{J}}^{\,(p)}_{\star,\,\leq 2r_{\rm opt,\star}}\right),
\end{equation}
where $\hat{\mathbf{J}}^{\,(p)}$ denotes the corresponding unit angular momentum vector projected onto plane $p$. In these 2D tests, class changes are similarly uncommon and occur exclusively in predictable geometric configurations. In the $xy$ projection, only LG1--gx81 changes class because one of its angular momentum vectors lies nearly along the line of sight. Its projected component is therefore small, making the inferred $\Delta\mathrm{PA}_{\rm 2D}$ sensitive to small transverse variations. The three $xz$ and four $yz$ changes similarly arise from small projected vectors or systems near the adopted thresholds.

\begin{figure}[tbp] 
\centering
\includegraphics[width=\columnwidth]{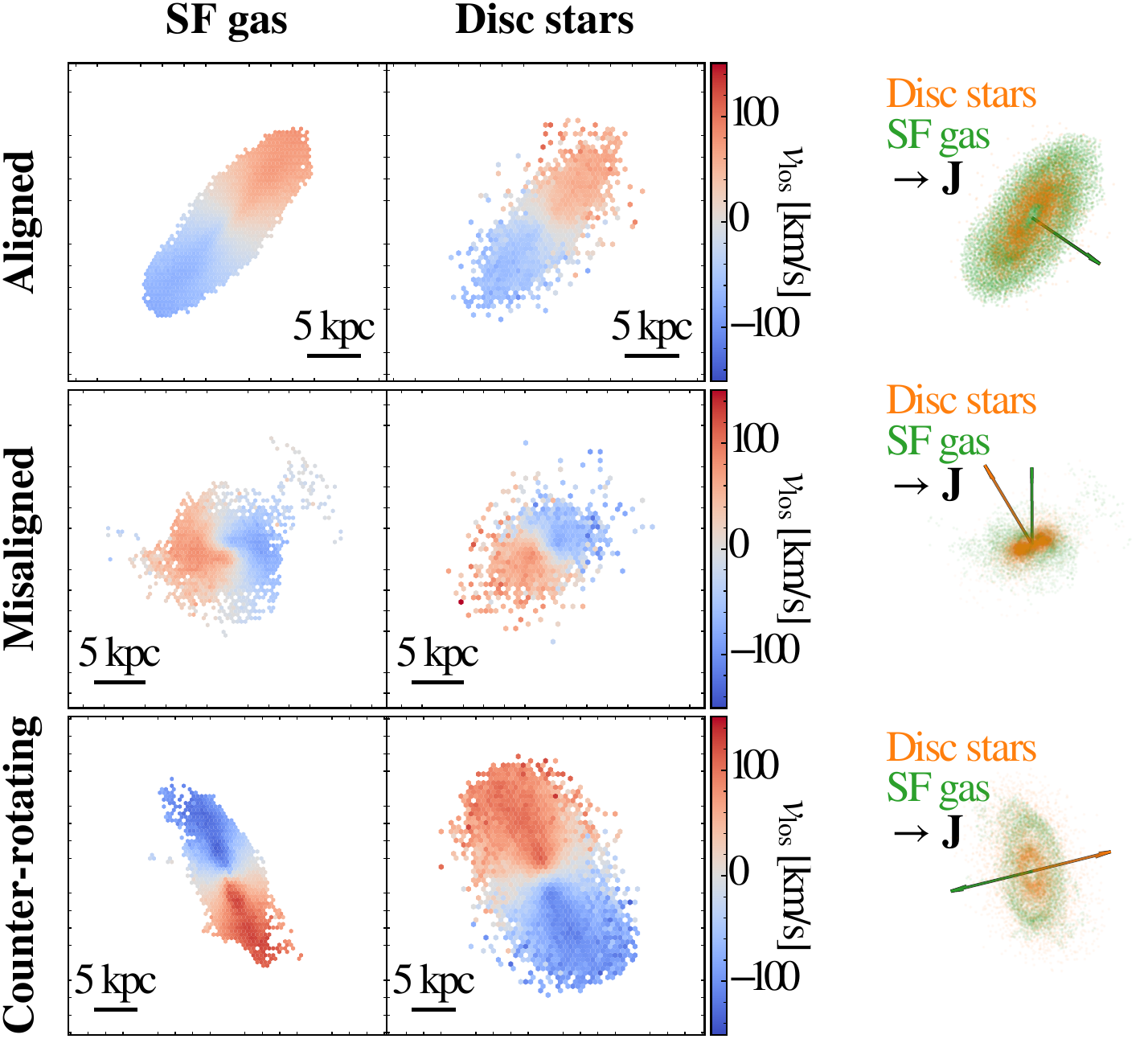}
\caption{Examples of aligned, misaligned, and counter-rotating galaxies at $z=0$. From top to bottom: LG1--gx2298, LG1--gx2097, and P4--gx428. In each row, the left and middle panels show the line-of-sight velocity fields ($v_{\rm los}$) of the SF gas and disc-star components, respectively, viewed along a random line of sight. The right panels show the corresponding 3D spatial distributions (disc stars in orange, SF gas in green). The arrows indicate their intrinsic 3D total angular momentum vectors ($\mathbf{J}$).}
\label{fig:pa_examples}
\end{figure}

Appendix~\ref{app:sample_overview} summarises the galaxy by galaxy comparison, examines the stability of the angular momentum vectors, and tests the effect of including all gas phases. Because $\psi$ is independent of viewing geometry (unlike projected offsets) and does not rely on a stellar decomposition (unlike disc-based definitions), it provides a highly stable fiducial baseline for our analysis. Finally, visual inspection of the mock velocity maps and 3D distributions for all sample galaxies confirmed that the measured angular offsets accurately reflect the true kinematic structure. Figure~\ref{fig:pa_examples} illustrates representative aligned, misaligned, and counter-rotating systems, comparing their line-of-sight velocity fields and intrinsic 3D orientations.

\subsection{Episode construction and transition bookkeeping}
\label{subsec:episodes}

For the time-resolved analysis up to $z=3.5$, we evaluate the kinematic history of each galaxy by applying the same sample selection criteria defined at $z=0$ (Sect.~\ref{subsec:sample}). Snapshots that do not satisfy these mass, particle count, and spatial centring requirements are flagged as ``not-resolved''. We retain for analysis only those portions of each history composed of ``resolved'' snapshots.

Accordingly, each resolved snapshot of each galaxy is classified into one of the three kinematic states: aligned, misaligned, or counter-rotating. We define a kinematic episode as any continuous time interval spent in a given state that contains at least one resolved snapshot. Intervals consisting entirely of not-resolved snapshots are discarded, and gaps of a single not-resolved snapshot do not by themselves induce a state transition. This bookkeeping provides the transition times and episode durations analysed throughout the paper.

We define an abrupt change as $|\Delta\psi|\geq35^\circ$ between consecutive snapshots. This amplitude can move a nearly aligned system beyond the $\psi=30^\circ$ alignment threshold within the typical snapshot spacing of $\simeq0.17$~Gyr. Changes separated by less than 0.35~Gyr (approximately two snapshots) are grouped into one event. Appendix~\ref{app:operational_sensitivity} tests the sensitivity to the angular threshold.

\begin{figure*}[tbp] 
\centering
\includegraphics[width=0.90\textwidth]{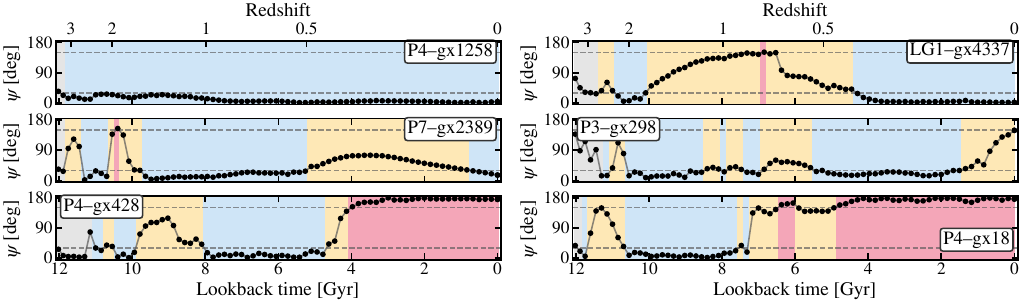}
\caption{Time evolution of the fiducial misalignment angle, $\psi$, for six representative galaxies. 
The panels show, from top left to bottom right, P4--gx1258, LG1--gx4337, P7--gx2389, P3--gx298, P4--gx428, and P4--gx18. 
The lower $x$-axis gives lookback time and the upper $x$-axis the corresponding redshift. 
Blue, yellow, and red shading mark aligned, misaligned, and counter-rotating intervals, respectively, using thresholds at $\psi=30^\circ$ and $\psi=150^\circ$. 
The grey band marks the initial interval whose onset lies outside the plotted range.
}
\label{fig:study_cases}
\end{figure*}

\subsection{Merger characterisation}
\label{subsec:merger_characterisation}

To interpret these evolutionary paths, we complement the episode analysis with a detailed catalogue of baryonic merger events for each central galaxy. These events trace the infall and coalescence of companion structures initially identified by the \textsc{subfind} algorithm \citep{Springel_2001,Dolag_2009}. Because the baseline merger trees linking these structures across time were generated via the AMIGA Halo Finder \citep[AHF;][]{Knollmann_2009}, temporary snapshot-to-snapshot tracking anomalies can occur. To ensure that only robust mergers are analysed, we applied a tailored cleaning procedure to remove spurious duplicate entries caused by this artificial fragmentation. Full details of the tree construction and cleaning are provided in \citet{CasanuevaVillarreal_2026_pipeline}. For each merger, we consider the baryonic mass ratio, $\mu_b$, the satellite gas fraction, $f_{\rm g}$, and the infall snapshot. Evaluated at infall, these quantities are
\begin{equation}
\mu_b \equiv
\frac{M_{\star,\mathrm{sat}} + M_{{\rm gas},\mathrm{sat}}}
     {M_{\star,\mathrm{cen}} + M_{{\rm gas},\mathrm{cen}}},
\qquad
f_{\rm g} \equiv
\frac{M_{{\rm gas},\mathrm{sat}}}
     {M_{\star,\mathrm{sat}} + M_{{\rm gas},\mathrm{sat}}},
\end{equation}
where ``sat'' and ``cen'' denote the satellite and central subhaloes. All stellar and gas masses are the total bound masses assigned by \textsc{subfind} at infall. We define infall at the crossing of the host virial radius ($R_{\rm vir}$). Across the 44-galaxy sample, we retain 395 mergers with $\mu_{\rm b}\geq10^{-3}$ and $z_{\rm infall}\leq3.5$. We divide them into very minor ($10^{-3}\leq\mu_{\rm b}<0.1$), minor ($0.1\leq\mu_{\rm b}<0.25$), and major ($\mu_{\rm b}\geq0.25$) mergers. The use of infall, first pericentre, and relative phase-space coordinates follows standard descriptions of satellite accretion orbits in cosmological simulations \citep[e.g.][]{Benson_2005,Wetzel_2011,Jiang_2015}.

For every merger, we measure the relative position ($\mathbf{r}_{\rm rel} \equiv \mathbf{r}_{\rm sat}-\mathbf{r}_{\rm cen}$) and velocity ($\mathbf{v}_{\rm rel} \equiv \mathbf{v}_{\rm sat}-\mathbf{v}_{\rm cen}$) of the satellite with respect to the central galaxy. From these, we define the orbital angular momentum vector $\mathbf{L}_{\rm orb} \equiv \mathbf{r}_{\rm rel}\times\mathbf{v}_{\rm rel}$, alongside standard orbital scalars: relative distance ($r_{\rm rel}$), relative speed ($v_{\rm rel}$), and the radial ($v_r$) and tangential ($v_t$) components of the relative velocity. We identify the first pericentric passage after infall, $t_{\rm peri1}$, from the first change in radial velocity from negative to non-negative. For short or truncated tracks, we use the first local minimum in $r_{\rm rel}$, or the global minimum when no local minimum is identified. Appendix~\ref{app:case_merger_support} gives the full implementation.

The stellar and SF gas angular momenta are computed from their particle distributions relative to the corresponding centre and bulk velocity. Thus, $\mathbf{L}_{\rm orb}$ describes the satellite orbit, whereas $\mathbf{J}_{\star}$ and $\mathbf{J}_{\rm SF}$ describe the net internal angular momenta of the baryonic components. The host stellar-to-SF gas angle at infall is $\psi$ evaluated at $t_{\rm infall}$.

\subsection{Gas-supply and origin diagnostics}
\label{subsec:gas_supply_origin}

To characterise the amount and origin of gas entering the central SF reservoir, we define the newly added SF gas at snapshot $i$ as the particles in the central SF component within $2\,r_{\rm opt,\star}$ at $i$ that were not in this component at $i-1$; its total mass is denoted by $M_{\rm new}(i)$. At the same snapshot, $M_{\rm pre-existing\,SF}(i)$ and $M_{\rm pre-existing\,gas}(i)$ are the remaining central SF and total gas masses within the same aperture after excluding the newly added particles. We define the impact of newly added SF gas as the percentage $100\,M_{\rm new}(i)/M_{\rm SF}(i)$, where $M_{\rm SF}(i)=M_{\rm new}(i)+M_{\rm pre-existing\,SF}(i)$.

We also measure the intrinsic 3D angular offset between the newly added SF gas and the pre-existing stellar component within the same aperture,
\begin{equation}
\theta_{\rm new,\star}(i)\equiv
\cos^{-1}\!\left(
\hat{\mathbf{J}}_{\rm new\,SF}(i)\cdot
\hat{\mathbf{J}}_{\star,\rm pre}(i)
\right),
\end{equation}
where $\hat{\mathbf{J}}_{\rm new\,SF}(i)$ is the total angular momentum vector of the newly added SF gas at snapshot $i$, and $\hat{\mathbf{J}}_{\star,\rm pre}(i)$ is the corresponding vector of the stellar component within $2\,r_{\rm opt,\star}$ formed before snapshot $i$. Throughout the rest of the paper, $\psi$ is reserved for the fiducial intrinsic 3D angle between the total central SF gas and the stellar component. Angles involving the newly added SF gas are denoted by $\theta$, with $\theta_{\rm new,\star}$ as the main quantity used in the paired and support diagnostics.

For each particle identified as newly added to the central SF component within $2\,r_{\rm opt,\star}$ at snapshot $i$, the immediate entry channel is assigned from its state at $i-1$. ``Cooling'' denotes non SF gas bound to the central subhalo and already within the aperture; ``infall'' denotes gas bound to the central subhalo but outside the aperture, irrespective of its SF state; and ``external'' denotes gas not associated with the central subhalo. Their mass fractions define $f_{\rm cooling}$, $f_{\rm infall}$, and $f_{\rm external}$, with $f_{\rm fresh}\equiv f_{\rm infall}+f_{\rm external}$.

To reconstruct the earlier history of this gas, we trace the same particles back to $z\approx7$ and classify their state before first entering the host $R_{\rm vir}$. Particles associated with mergers are assigned first to the ``merger'' class. Among the remaining particles, ``smooth accretion unbound'' denotes gas with no recovered substructure association; ``smooth accretion stripped'' denotes gas that is unbound in the snapshot immediately preceding its first $R_{\rm vir}$ entry but was associated with another substructure at an earlier snapshot; and ``stripped'' denotes gas that, in the snapshot immediately preceding its first entry into the host $R_{\rm vir}$, is bound to a substructure that does not merge with the host by $z=0$. The mass fractions in these origin classes are evaluated as functions of both the SF entry time, i.e. when particles first enter the central SF reservoir within $2\,r_{\rm opt,\star}$, and the time of first \textsc{subfind} association with the central subhalo.

At snapshot $i$, we classify each newly added particle according to its angular offset from the total angular momentum of the stellar component formed before $i$ within $2\,r_{\rm opt,\star}$ and from that of the remaining central gas after excluding the newly added particles.

We follow the subsequent evolution through the delay between SF entry and conversion into a star particle, considering particles that form stars by $z=0$. We also classify the $z=0$ phase and \textsc{subfind} association of all tracked particles. ``Central stars'' and ``central gas'' denote particles assigned to the central subhalo, without a radial aperture restriction; the ``central gas'' class includes all gas phases. ``Unbound gas'' denotes gas not assigned to any bound substructure. All particle based mass fractions described above are weighted by particle mass at SF entry. The delay fractions are normalised by the mass that forms stars by $z=0$, whereas the fate fractions are normalised by the total newly added SF gas mass.

\begin{figure*}
\centering
\includegraphics[width=0.95\textwidth]{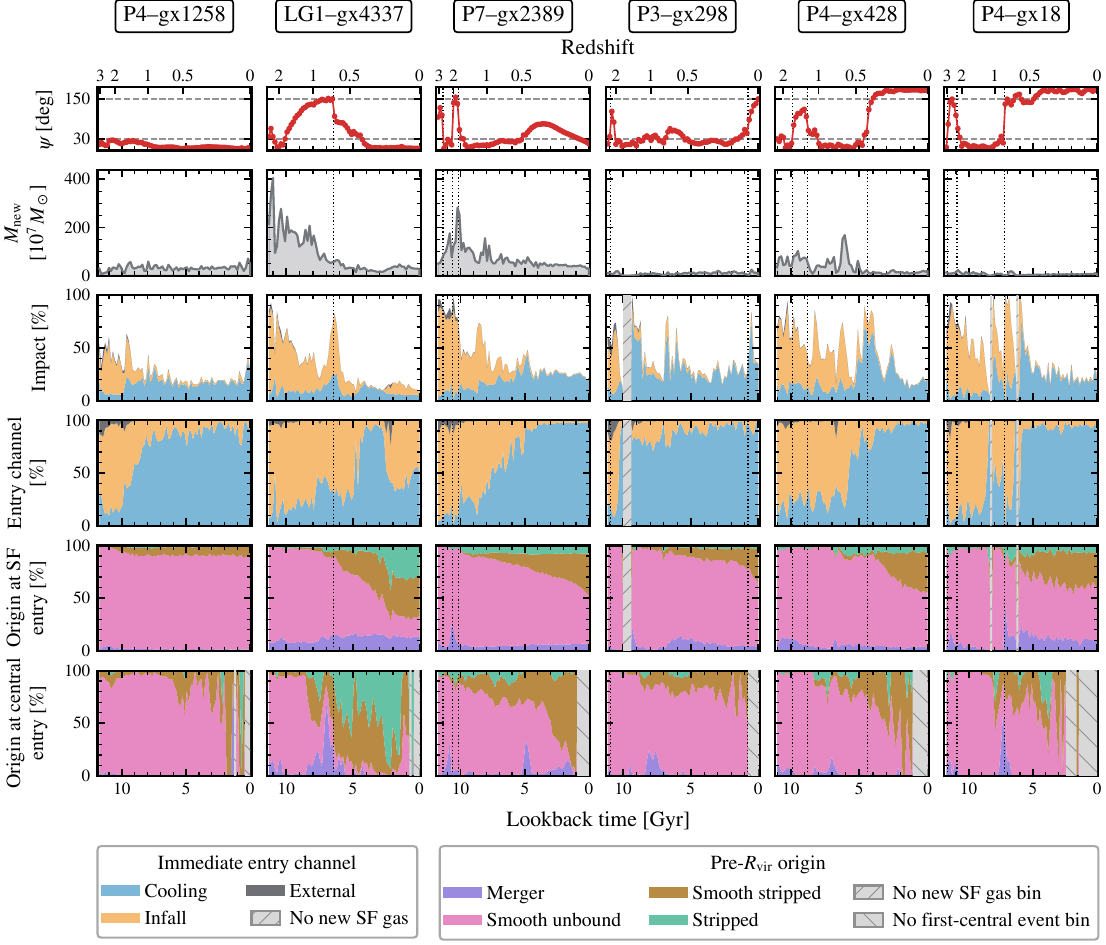}
\caption{Evolution of the gas--stellar angular offset, $\psi$, and of the mass, impact, immediate entry channel, and pre-$R_{\rm vir}$ origin of newly added SF gas in the six representative galaxies. Newly added SF gas comprises particles in the central SF component within $2\,r_{\rm opt,\star}$ at snapshot $i$ that were not in this component at $i-1$. Columns show, from left to right, P4--gx1258, LG1--gx4337, P7--gx2389, P3--gx298, P4--gx428, and P4--gx18, with lookback time along the lower axis and redshift along the upper axis. Rows show, from top to bottom, $\psi$; $M_{\rm new}$; the impact of newly added SF gas, $100\,M_{\rm new}(i)/M_{\rm SF}(i)$, decomposed by immediate entry channel; the cooling, infall, and external fractions; the pre-$R_{\rm vir}$ origin classes evaluated at SF entry; and the same particles evaluated at their first \textsc{subfind} association with the central subhalo. Entry channels and pre-$R_{\rm vir}$ origin classes are defined in Sect.~\ref{subsec:gas_supply_origin}. Rows 3--6 are cumulative stacked area plots, with each contribution given by the vertical thickness of its layer. Dotted lines mark $|\Delta\psi|\geq35^\circ$. Hatching marks snapshots with no newly added SF gas in rows 3--5. In the bottom row, it marks time bins containing no particles whose first association with the central subhalo occurs at that time.}
\label{fig:case6_supply_origin}
\end{figure*}

\begin{figure*}
\centering
\includegraphics[width=0.95\textwidth]{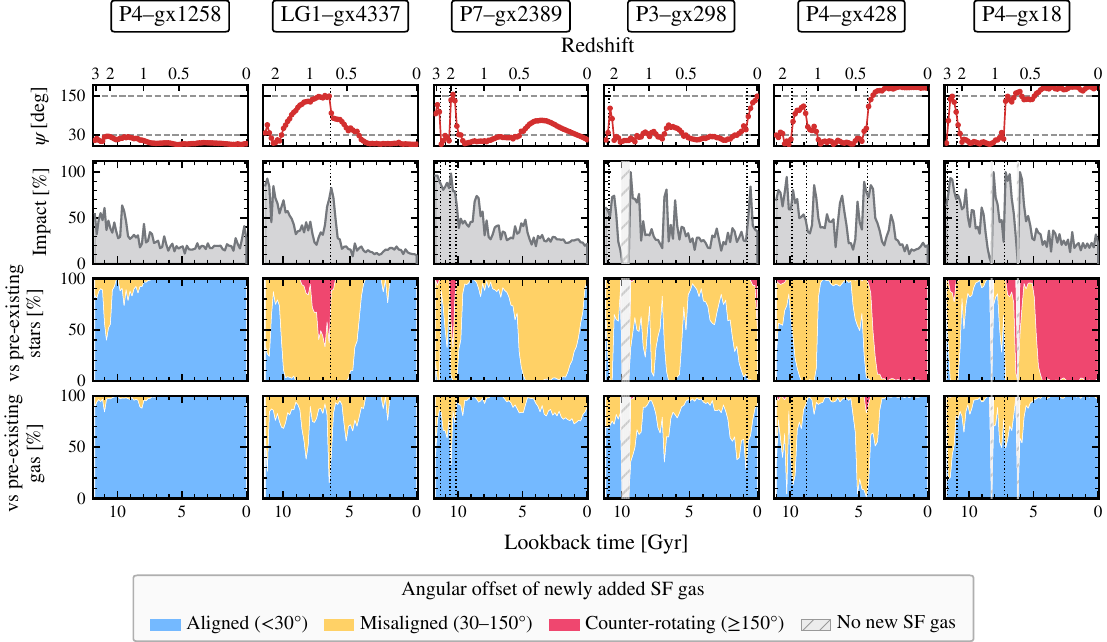}
\caption{Angular orientation of newly added SF gas in the six representative galaxies. The columns and time axes follow Fig.~\ref{fig:case6_supply_origin}. Rows show, from top to bottom, $\psi$; the impact of newly added SF gas; and the mass fractions of newly added particles classified as aligned
($<30^\circ$), misaligned ($30^\circ$--$150^\circ$), or counter-rotating ($\geq150^\circ$) relative to the pre-existing stellar component and central gas reservoir, respectively. At each snapshot, the pre-existing stellar component comprises stars within $2\,r_{\rm opt,\star}$ formed before that snapshot, whereas the pre-existing central gas reservoir comprises all gas within the same aperture after excluding the newly added SF gas. Dotted lines mark $|\Delta\psi|\geq35^\circ$. Hatching marks snapshots with no newly added SF gas.}
\label{fig:case6_pa_context}
\end{figure*}

\section{Six representative systems}
\label{sec:case_pathways}

While the $z=0$ kinematic states describe the present-day relative orientations, they provide limited insight into the physical mechanisms that establish these configurations. In order to understand how galaxies build and reorient their angular momentum, we must trace their evolution over cosmic time. In this section, we analyse the detailed histories of six selected systems: three aligned at $z=0$ (LG1--gx4337, P4--gx1258, P7--gx2389), one misaligned (P3--gx298), and two counter-rotating (P4--gx428, P4--gx18). These cases were chosen to illustrate the five distinct evolutionary pathways identified in our episode catalogue: continuous aligned evolution, temporary misalignment followed by recovery, a transitional near-counter-rotating phase, and two counter-rotating channels. The frequency and duration of these evolutionary behaviours across the full sample are quantified later in Sect.~\ref{sec:global_stats}.

Figures~\ref{fig:study_cases}--\ref{fig:case6_delay_fate} link the kinematic histories of these six systems to the mass, origin, angular momentum, and fate of gas entering the central SF reservoir.
  
\subsection{Physical drivers of kinematic transitions}
\label{sec:case_mergers}

As we will demonstrate across these examples, the recurring drivers of kinematic transitions are not merely the occurrence of mergers, but the timing, orientation, and the fractional contribution of newly added SF gas to the total central SF reservoir, coupled with the response of the pre-existing reservoir. High-impact gas additions can either sustain a non-aligned state or be followed by a full kinematic recovery, depending strictly on how the incoming material mixes with the gas already in place. To interpret these events physically, we trace the complete origin of the incoming gas (whether it arrives via smooth accretion, stripped material, or mergers; Sect.~\ref{subsec:gas_supply_origin}) alongside the detailed orbital and baryonic kinematics of the mergers themselves (Sect.~\ref{subsec:merger_characterisation}).

\paragraph{The hidden complexity behind present-day classes.}
Consistent with previous simulation work, we find that the $z=0$ kinematic class does not uniquely specify the preceding history \citep{Casanueva_2022,Baker_2025a}. As shown by the time evolution of $\psi$ (Fig.~\ref{fig:study_cases}), aligned galaxies within this sample include both the kinematically undisturbed reference case (P4--gx1258) and systems that recover alignment after substantial earlier non-aligned intervals (LG1--gx4337 and P7--gx2389). Likewise, the two counter-rotating end points are reached through different histories: P4--gx428 through repeated kinematic transitions culminating in a late-time counter-rotation, and P4--gx18 through an earlier major transition that establishes a prolonged counter-rotating episode. The relevant distinction is therefore not only the final class, but also the number, duration, and ordering of the episodes that lead to it.

\begin{figure*}
\centering
\includegraphics[width=0.95\textwidth]{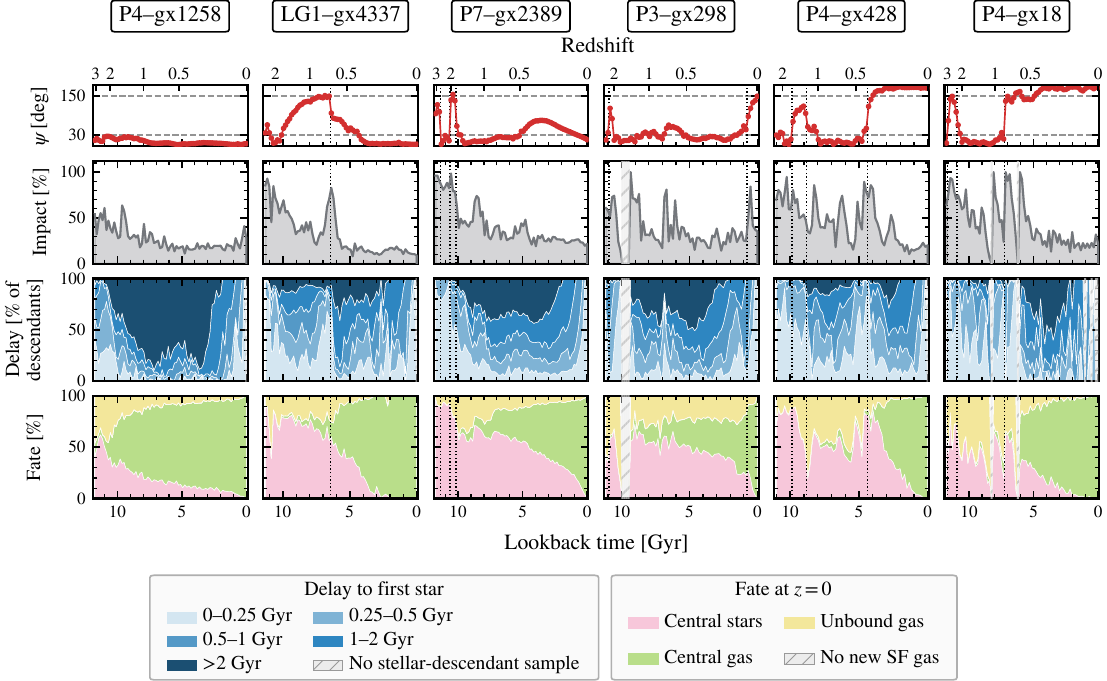}
\caption{Star-formation delay and $z=0$ fate of newly added SF gas in the six representative galaxies. The columns and time axes follow Fig.~\ref{fig:case6_supply_origin}. Rows show, from top to bottom, $\psi$; the impact of newly added SF gas; the delay between entry into the central SF reservoir and conversion into a star particle for particles that form stars by $z=0$; and the $z=0$ fate of all newly added SF gas. The coloured layers in the third and bottom rows show, respectively, the fractions of particles in each delay time bin among those that form stars by $z=0$ and the $z=0$ fate fractions of all newly added SF gas. ``Central stars'' and ``central gas'' denote particles that are stars and gas, respectively, at $z=0$ and are assigned by \textsc{subfind} to the central subhalo; the central gas class includes all gas phases. ``Unbound gas'' denotes gas not assigned to any bound substructure. Other outcomes contribute less than 0.3\% in every example. Dotted lines mark $|\Delta\psi|\geq35^\circ$. Hatching in the third row marks snapshots with no particles that form stars by $z=0$, while hatching in the bottom row marks snapshots with no newly added SF gas.}
\label{fig:case6_delay_fate}
\end{figure*}

\paragraph{The apparent interplay between incoming gas and reservoir response.} Fig.~\ref{fig:case6_supply_origin} separates the immediate entry channel of newly added SF gas from its pre-$R_{\rm vir}$ origin. The two lower origin panels place the same particles at different stages of their evolution: SF entry in the fifth row and their first \textsc{subfind} association with the central subhalo in the sixth row, i.e. the first snapshot in which they are gravitationally bound to the central subhalo. Across the six systems, the time-integrated supply is dominated by smooth accretion unbound, while the merger-origin contribution spans 1.7\% in P4--gx1258 to 10.1\% in LG1--gx4337. In LG1--gx4337, first central association bins at $t_{\rm lb}=6.85$--$7.01$~Gyr contain 54.4--70.5\% merger-origin gas and precede the abrupt change at $t_{\rm lb}\simeq6.38$~Gyr. P7--gx2389 shows a corresponding first central association fraction of 33.1\% at $t_{\rm lb}=10.85$~Gyr, followed by a 26.6\% merger-origin fraction when the gas enters the central SF reservoir at $t_{\rm lb}=10.70$~Gyr. In P4--gx18, merger-origin gas accounts for 46.0\% of the newly added SF gas at $t_{\rm lb}=7.16$~Gyr, when $\psi$ changes from $17.6^\circ$ to $134.4^\circ$. One snapshot earlier, merger-origin particles account for 60.7\% of the mass first associated with the central subhalo. The remaining systems do not show comparably concentrated merger-origin peaks. In several cases, merger-origin gas is temporally associated with abrupt changes either when it first becomes bound to the central subhalo or when it subsequently joins the central SF reservoir.

Figure~\ref{fig:case6_pa_context} shows that the response to high-impact additions depends on their angular offset relative to the pre-existing components. P4--gx1258 remains aligned despite high-impact additions and transiently elevated angular offsets of the newly added SF gas; its total SF gas--stellar angle remains below $30^\circ$. By contrast, P4--gx428 and P4--gx18 undergo repeated or high-impact misaligned additions during their evolution towards long-lived counter-rotation, whereas LG1--gx4337 and P7--gx2389 later recover alignment as subsequent gas supply no longer maintains the non-aligned configuration. The final kinematic state therefore depends on the mass and angular momentum coupled to the pre-existing gas reservoir, not on a single entry channel or pre-$R_{\rm vir}$ origin.
  
\paragraph{Mergers appear as conditional triggers rather than one-to-one causes.}
In these examples, mergers occur near episode boundaries, but neither merger occurrence nor $\mu_{\rm b}$ alone determines the subsequent kinematic evolution. As detailed in Appendix~\ref{app:case_merger_support}, which also defines the orbital classes used below, encounters with comparable mass ratios and orbital configurations precede different evolutionary paths. P3--gx298 experiences a coplanar-retrograde, radial-dominated merger with $\mu_{\rm b}=0.20$ and remains in a prolonged transitional state, whereas comparable encounters in P4--gx428 ($\mu_{\rm b}=0.22$) and P4--gx18 ($\mu_{\rm b}=0.25$) precede long-lived counter-rotation. Likewise, polar or inclined mergers precede both recovery in LG1--gx4337 and P7--gx2389 and prolonged non-alignment in P3--gx298.

Furthermore, merger-linked material is not deposited instantaneously. Using lookback times, we define $\Delta t_{90} \equiv t_{\star,90}-t_{{\rm SF},90}$, where $t_{\star,90}$ and $t_{{\rm SF},90}$ mark when the satellite stellar and SF gas components reach 90\% of their final $z=0$ contributions to the central \textsc{subfind} galaxy. For the key events in Table~\ref{tab:case_merger_diagnostics}, $\Delta t_{90}$ spans from $\simeq1.5$ to $10.6$~Gyr. This wide margin confirms that mergers supply and process material on a timescale that is largely decoupled from the immediate kinematic response of the central gas reservoir. While mergers can provide conditions for reorientation, the subsequent evolutionary path depends on the delivery of their material and on its coupling to the host gas reservoir; orbital geometry alone does not determine the outcome.

\paragraph{Persistence and recovery are governed by the host gas reservoir.}
Figure~\ref{fig:case6_delay_fate} shows the $z=0$ phase and subhalo association of particles identified as newly added SF gas at each snapshot in the six representative galaxies (Sect.~\ref{subsec:gas_supply_origin}). In the long-lived counter-rotating intervals of P4--gx18 and P4--gx428, 64\% and 55\% of the newly added gas survives as central gas at $z=0$, compared to only 24\% and 40\% converting to stars, respectively. For the fraction that does form stars, the median delay is short ($\simeq0.3$~Gyr), indicating that persistence is driven by the survival of a dynamically important, misaligned gas reservoir rather than by inefficient star formation.

Conversely, recovering systems deplete or dilute the tilted reservoir. In LG1--gx4337, the newly added SF gas during the main non-aligned episode is largely removed from the central gas phase by $z=0$: 83\% is converted into central stars and only 9\%
remains as central gas. P7--gx2389 retains more gas (39\%), but subsequent accretion overwhelms the tilted component, driving the
system back to alignment. Consistent with the merger transfer diagnostics (Appendix Fig.~\ref{fig:phase_handoff_cases}), the stellar and gaseous components of the same accretion event settle at different times. Consequently, across all events analysed here the stellar and SF gas components initially associated with the same merger reach their retained $z=0$ contributions to the central subhalo at different times.

\paragraph{Summary of the case studies.}
Overall, these six representative systems show that gas--stellar decoupling is governed by the competition between incoming supply and the central reservoir. This aligns with previous work identifying mergers as conditional triggers rather than the sole channel for reorientation \citep{Lagos_2015, Khim_2020a, Casanueva_2022}, and with recent simulations showing that the persistence of non-alignment depends on gas replacement, cancellation, and depletion, rather than on external perturbations alone \citep{Starkenburg_2019, Cenci_2023, Han_2024, Peirani_2025, Baker_2025a}. Guided by this physical framework, we establish the terminology used in the subsequent statistical analysis: the onset of non-alignment, recovery when later supply stops reinforcing the tilt, persistence through misaligned replenishment, reservoir replacement, and the delayed conversion of newly accreted gas into stars.

\begin{figure*}[tbp] 
\centering
\includegraphics[width=0.9\textwidth]{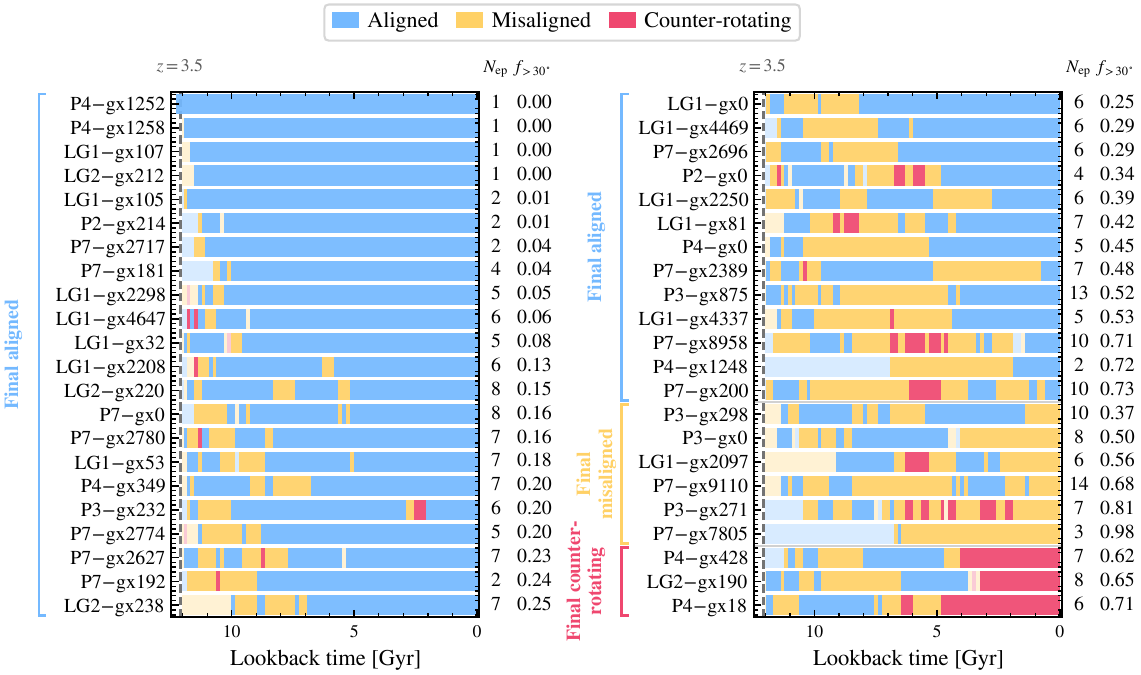}
\caption{Time-history map for the 44 galaxies used in the statistical analysis. Rows are grouped by the final $z=0$ class and, within each group, ordered by the fraction of valid tracked time spent at $\psi>30^\circ$. Blue, orange, and red segments indicate aligned, misaligned, and counter-rotating states, respectively. Saturated segments denote the numerically resolved snapshots used for episode-based statistics, while pale segments are shown only to provide visual continuity for intervals that fall below our numerical quality thresholds. The dashed vertical line marks the adopted $z=3.5$ boundary. The right-hand columns list the number of counted episodes, $N_{\rm ep}$, and the fraction of valid tracked time spent above $30^\circ$, $f_{>30^\circ}$. Episodes are counted following the definition in Sect.~\ref{subsec:episodes}, and since misaligned and counter-rotating are both non-aligned states, a transition between them does not by itself start a new episode; a counter-rotating interval nested within an otherwise misaligned stretch is therefore counted as part of the same episode.}
\label{fig:history_atlas}
\end{figure*}

\begin{figure*}[tbp] 
\centering
\includegraphics[width=0.9\textwidth]{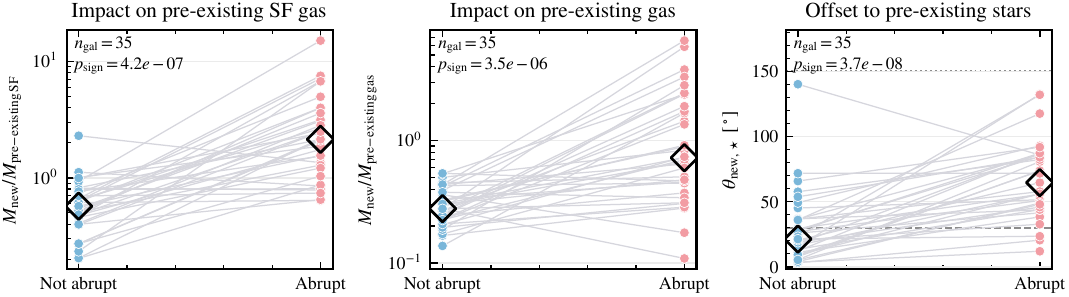}
\caption{Galaxy-level paired contrasts between abrupt-transition intervals and the complementary non-abrupt intervals. For each galaxy, coloured points show the within-galaxy medians and grey segments join the two conditions; open diamonds mark the sample medians. From left to right, the diagnostics are the mass ratios of newly added SF gas to the pre-existing SF and total gas reservoirs, respectively, and the angular offset between the newly added SF gas and the pre-existing stellar component. The paired sample contains 35 galaxies.}
\label{fig:paired_signal}
\end{figure*}

\section{Global statistics for the 44-galaxy sample}
\label{sec:global_stats}

We now test whether the physical mechanisms identified in the six representative systems hold generally across the full 44-galaxy sample.

\subsection{Temporal pathways and paired abrupt-interval statistics}
\label{subsec:results_stats}

To test whether kinematic non-alignment is driven by misaligned gas additions that achieve dynamical importance, we compare the conditions during abrupt-transition intervals against complementary non-abrupt intervals within each galaxy. This paired statistical design isolates internal evolution by suppressing galaxy-to-galaxy variations in mass, environment, and baseline orientation, yielding robust time-averaged contrasts over the $z=3.5$ to $z=0$ tracking window.

For each diagnostic, we compute a paired shift for each galaxy, defined as the difference between its median diagnostic value during abrupt-transition intervals and its corresponding median during non-abrupt intervals. We then report the sample-wide median of these paired shifts. Our statistical evaluation is divided into two tiers. Galaxy-level comparisons treat each galaxy as a single independent pair, which we evaluate using two-sided sign tests and 95\% bootstrap confidence intervals (CIs). Event-level comparisons instead pool individual occurrences across the entire sample (e.g., all mergers near versus far from boundaries) and are evaluated using permutation tests. The resulting paired signal is highly robust to the adopted alignment and abruptness thresholds (Appendix~\ref{app:operational_sensitivity}).

Figure~\ref{fig:history_atlas} shows the reconstructed histories for the full sample and confirms that the present-day class captures only part of the earlier history. Of the 35 galaxies aligned at $z=0$, only 4 remain aligned throughout the tracked interval; the other 31 experience at least one earlier non-aligned episode. Across the full sample, 40 galaxies enter a misaligned or counter-rotating state at some point, and 31 of those 40 later realign by $z=0$, making present-day alignment often a recovered state. Appendix~\ref{app:temporal_summary} summarises the corresponding history classes and makes the episode duration distributions explicit.

Non-aligned episodes span a broad duration distribution (Appendix Fig.~\ref{fig:episode_duration_distributions}). Of these, 36\%, 19\%, and 7\% last at least 1, 2, and 5~Gyr, respectively. The pooled medians are 0.61~Gyr for non-aligned intervals and 0.76~Gyr for aligned ones. Among the 40 galaxies containing at least one aligned and one non-aligned episode, the within-galaxy comparison yields a median paired shift of $-0.08$~Gyr, with a $p$-value of 0.87 (hereafter $p$) from a two-sided sign test, confirming that the two duration distributions are not statistically distinct. Long-lived non-alignment is therefore best interpreted as the tail of a broad distribution rather than as a separate temporal population.

The paired comparison in Fig.~\ref{fig:paired_signal} is restricted to galaxies that enter both temporal regimes: abrupt-transition and non-abrupt intervals. Under the fiducial threshold ($|\Delta\psi|\geq35^\circ$), 35 of the 44 galaxies contain both types of intervals with valid diagnostic measurements. The nine excluded systems are all aligned at $z=0$ and never enter the abrupt regime; relative to the paired subset, they have fewer mergers, lower maximum baryonic mass ratios, and much less tracked non-aligned time. These exclusions therefore identify dynamically quieter galaxies, not systems entirely devoid of accretion or non-aligned behaviour. P4--gx1248 is the clearest counterexample, with a long non-aligned interval but no individual step that satisfies the abrupt threshold. Appendix~\ref{app:paired_subset_eligibility} summarises the corresponding eligibility breakdown.

Figure~\ref{fig:paired_signal} shows the main full-sample signal. Within the paired subset of 35 galaxies, all three diagnostics shift in the same direction. The mass ratio of newly added to pre-existing SF gas rises from a median of 0.57 to 2.14 when normalised by the pre-existing SF reservoir ($M_{\rm new}/M_{\rm pre-existing\,SF}$), and from 0.28 to 0.72 when normalised by the pre-existing total gas reservoir ($M_{\rm new}/M_{\rm pre-existing\,gas}$). The median paired shifts are $+1.82$ ($95\%$ CI $[0.78,\,2.42]$, $p=4.2\times10^{-7}$) and $+0.42$ ($95\%$ CI $[0.13,\,1.17]$, $p=3.5\times10^{-6}$), respectively. Concurrently, the angular offset between the newly added SF gas and the pre-existing stars ($\theta_{\rm new,\star}$) increases from a median of $21.2^\circ$ in non-abrupt intervals to $64.6^\circ$ in abrupt-transition intervals, yielding a median paired shift of $+30.5^\circ$ ($95\%$ CI $[24.7^\circ,\,46.8^\circ]$, $p=3.7\times10^{-8}$). In summary, the median paired shifts are positive and highly significant across all three diagnostics.

These robust global statistics confirm an internal evolutionary link: across the sample, intervals close to abrupt changes are precisely those where the newly added SF gas becomes more dynamically important relative to the pre-existing reservoir, and more misaligned with the pre-existing stars.

\subsection{Fresh supply, origin, and reservoir response}
\label{subsec:conditional_mergers}

The paired abrupt/not-abrupt contrasts identify the local conditions associated with rapid reorientation, but not the provenance of the incoming gas or its subsequent evolution. We therefore examine the immediate entry channel and pre-$R_{\rm vir}$ origin of these particles (as described in Sect.~\ref{subsec:gas_supply_origin}), alongside their post-entry response and the merger state of the host.

\begin{figure}[tbp] 
\centering
\includegraphics[width=0.85\columnwidth]{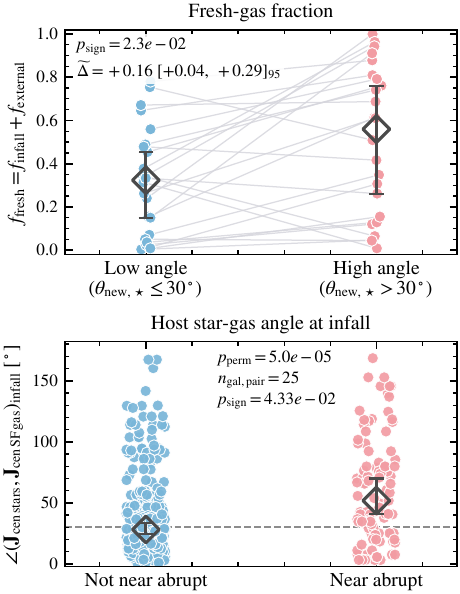}
\caption{Support diagnostics for abrupt transitions. Upper panel: paired comparison of fresh gas fraction between low-angle ($\theta_{\rm new,\star}\leq30^\circ$) and high-angle ($\theta_{\rm new,\star}>30^\circ$) additions of SF gas relative to the pre-existing stars. Coloured points and grey segments show within-galaxy medians. Lower panel: angle between the host stellar and SF gas angular momentum vectors at infall for the full merger catalogue, divided according to whether an abrupt event ($|\Delta\psi|\geq35^\circ$) lies within 0.35~Gyr of infall. The remaining mergers have larger separations or belong to galaxies without an abrupt event. Coloured points show the 389 mergers with measurable angles, and the dashed line marks the $30^\circ$ alignment threshold. In both panels, open diamonds and whiskers give the corresponding medians and 95\% bootstrap confidence intervals. The upper-panel two-sided paired sign test gives $n=24$ and $p=2.3\times10^{-2}$; $\widetilde{\Delta}$ denotes the median paired shift. The lower-panel two-sided event permutation test gives $p_{\rm perm}=5.0\times10^{-5}$; the two-sided paired-galaxy sign test gives $n=25$ and $p=4.3\times10^{-2}$.}

\label{fig:support_signal}
\end{figure}

The upper panel of Fig.~\ref{fig:support_signal} compares the fresh-gas fraction between low- and high-angle additions of newly incorporated SF gas, measured relative to the pre-existing stellar component. This comparison is available for 24 galaxies, because it requires particle-origin classifications and valid angular offset measurements in both angle regimes; this sub-sample remains kinematically representative of the full population (Appendix~\ref{app:paired_subset_eligibility}).
Among them, the median fresh fraction increases from 0.32 to 0.56 between the low-angle ($\theta_{\rm new,\star}\leq30^\circ$) and high-angle ($\theta_{\rm new,\star}>30^\circ$) contexts, with a median paired shift of $+0.155$ ($95\%$ CI $[0.043,\,0.294]$, $p=2.3\times10^{-2}$). High-angle intervals are therefore preferentially associated with gas that has recently entered through infall or external transfer. A redshift-split extension of the immediate-entry analysis is presented in Appendix~\ref{app:origin_redshift}.

The same particle sample reveals an origin-dependent angular hierarchy. Relative to the pre-existing gas reservoir, merger-origin gas is consistently the most tilted accretion channel. Defining $z_{\rm detected}$ as the redshift when a particle first joins the central SF reservoir, the pooled distribution of individual particle offsets at $z_{\rm detected} \geq 2$ yields median values of $38.4^\circ$, $33.0^\circ$, and $34.2^\circ$ for merger-origin, smooth accretion unbound, and smooth accretion stripped gas, respectively. This hierarchy persists at $z_{\rm detected} < 2$, albeit with smaller absolute offsets. By contrast, the distinction between channels weakens when measured relative to the pre-existing stellar component, and becomes heavily diluted if the angles are evaluated at initial $R_{\rm vir}$ crossing instead of at central SF entry. 

Consequently, the origin-dependent kinematic separation is established predominantly at the point where the gas joins the central SF reservoir, rather than at its initial entry into the halo.

The paired aligned$\rightarrow$misaligned/misaligned$\rightarrow$aligned contrasts reveal that post-entry evolution differs between the two transition directions. Among the 31 galaxies with both aligned$\rightarrow$misaligned and misaligned$\rightarrow$aligned transitions, the median replacement timescale of the central SF reservoir, $\tau_{\rm repl}$, is 0.30~Gyr for aligned$\rightarrow$misaligned transitions and 0.52~Gyr for misaligned$\rightarrow$aligned ones, with a median paired difference of $+0.18$~Gyr ($95\%$ CI $[0.15,\,0.25]$, $p=3.4\times10^{-5}$). The descendant-fate split is weaker but consistent with the same pattern: the median fraction of newly added gas remaining in the central gas phase is 0.10 for aligned$\rightarrow$misaligned transitions and 0.13 for misaligned$\rightarrow$aligned ones, with a median paired difference of $+0.008$ ($95\%$ CI $[0.003,\,0.032]$, $p=1.1\times10^{-2}$). By contrast, the star-formation delay bins do not show a significant directional separation, and the duration of the following episode is likewise indistinguishable ($p=0.47$). The global asymmetry, therefore, lies in the replacement and retention of the central SF reservoir rather than in a shift of the delay distribution.

The lower panel of Fig.~\ref{fig:support_signal} compares the angle between the host stellar and SF gas angular momentum vectors at infall for the full merger catalogue. We identify mergers for which infall and the nearest abrupt event ($|\Delta\psi|\geq35^\circ$) are separated by at most 0.35~Gyr. The remaining mergers either have larger separations or occur in galaxies without an abrupt event. This interval spans approximately two snapshot intervals and matches the event-grouping interval in Sect.~\ref{subsec:episodes}. Among the 389 mergers with a measurable host angle, the median is $51.8^\circ$ for mergers within this interval and $28.2^\circ$ for the remaining sample, a difference of $+23.6^\circ$ ($p_{\rm perm}=5.0\times10^{-5}$). Among the 25 galaxies containing both types of merger, the median within-galaxy difference is $+25.5^\circ$ ($95\%$ CI $[4.9^\circ,\,32.6^\circ]$, $p=4.3\times10^{-2}$). The host angle at infall is therefore larger for mergers temporally close to abrupt events. Appendix~\ref{app:merger_abrupt_distance} shows the corresponding separation distribution for mergers in galaxies with at least one abrupt event.

At the fiducial 0.35~Gyr window, no significant timing excess is detected for very minor mergers. Minor and major mergers show an excess around abrupt events at both infall and first pericentre. When all transitions between aligned and non-aligned episodes are included, the excess remains at both milestones for minor mergers and only at first pericentre for major mergers. Appendix~\ref{app:merger_linked_boundaries} gives the null-model definition, significance values, and window dependence.

Taken together, Figs.~\ref{fig:paired_signal} and~\ref{fig:support_signal} show that abrupt changes in $\psi$ occur when newly added SF gas becomes dynamically important relative to the pre-existing SF gas, when it is strongly tilted relative to the pre-existing stars, and when it is preferentially supplied by fresh infall or external transfer. Subsequent evolution depends on whether central SF gas is replaced or retained. Minor and major mergers occur in excess near abrupt events at infall and first pericentre. Those close to abrupt events enter hosts with larger angles between stellar and SF gas angular momenta at infall. The dominant mechanism is therefore competition between pre-existing and newly supplied gas, with mergers providing one channel for establishing or reinforcing it. Appendix~\ref{app:stat_extensions} collects robustness tests, supporting diagnostics, and counterexamples.

\section{Discussion}
\label{sec:discussion}

This work adds two main elements to the existing literature. First, we extend single-epoch demographic surveys and sample-averaged simulation studies by reconstructing continuous kinematic pathways within individual galaxies. We evaluate the associated physics using paired statistical contrasts to compare each galaxy directly against its own internal baseline. Second, we explicitly separate the immediate entry channel into the central SF reservoir, the pre-$R_{\rm vir}$ origin of the same particles, and the post-entry response of the central reservoir. This distinction isolates the origin of the gas, the timing of its dynamical impact, and its coupling to the material already in place.

Building on this framework, our results link abrupt reorientation to intervals in which newly added SF gas enters at a large angular offset and becomes dynamically important relative to the pre-existing central reservoir. Our temporal census places this mechanism in a broader evolutionary context. We find that many galaxies aligned at $z=0$ underwent one or more earlier non-aligned phases before ultimately realigning. The corresponding duration statistics confirm that long-lived non-alignment represents the extreme tail of a broad, continuous duration distribution, representing transient evolutionary phases rather than a kinematically distinct population.

Aligned$\rightarrow$misaligned transitions are associated with shorter replacement timescales than misaligned$\rightarrow$aligned ones, yet the star-formation delay distributions show no comparable directional separation. The kinematic outcome is therefore governed by the evolution of the central SF reservoir: non-alignment persists while tilted gas dominates its angular momentum. Alignment is recovered when this gas is consumed by star formation, removed from the central SF reservoir, has its angular momentum redistributed and partially cancelled through interactions with differently oriented gas, or is outweighed by subsequently supplied gas aligned with the stellar component \citep{Casanueva_2022,Cenci_2023,Peirani_2025,Baker_2025a}.
  
To quantify how rapidly the gas--stellar kinematic state changes, we calculate the finite-difference rate $|\psi_{i+1}-\psi_i|/|t_{i+1}-t_i|$ between consecutive valid snapshots. Intervals whose endpoints belong to different kinematic classes have a median rate of $115^\circ\,{\rm Gyr}^{-1}$; these include crossings of the $30^\circ$ or $150^\circ$ boundaries with $|\Delta\psi|<35^\circ$. Intervals satisfying the fiducial abrupt criterion, $|\Delta\psi|\geq35^\circ$, have a median rate of $406^\circ\,{\rm Gyr}^{-1}$. These intervals therefore have a substantially higher median rate of change in the relative gas--stellar orientation than intervals selected by a change of kinematic class.

The origin diagnostics confirm that this reservoir competition involves all three accretion categories: merger, smooth, and stripped channels. Descriptively, the pooled redshift-split measurements indicate that merger-origin gas is typically the most tilted component at central SF entry, especially relative to the pre-existing gas, while both smooth and stripped channels still contribute substantially. Importantly, the kinematic separation between these origin classes is much clearer at central SF entry than at first $R_{\rm vir}$ crossing. By the time gas first enters the halo, all channels are already broadly tilted and weakly separated. The decisive decoupling step therefore occurs during the subsequent delivery of gas into the central SF reservoir and its interaction with the material already present there \citep[e.g.,][]{Casanueva_2022, Cenci_2023, Peirani_2025}. In that sense, our results help reconcile the long-standing tension between merger-centred and smooth-accretion interpretations \citep{Davis_2011,Lagos_2015}: all supply channels contribute meaningfully, while the ultimate kinematic outcome remains governed by the central reservoir response \citep[see also][]{Baker_2025a}.

The timing analysis in Appendix~\ref{app:merger_linked_boundaries} shows that minor and major mergers are overrepresented around abrupt transitions at infall and first pericentre. Mergers close to these transitions enter hosts with larger angles between stellar and SF gas angular momenta at infall, indicating stronger pre-existing decoupling. Both mass classes are associated with transitions towards alignment and non-alignment, with no systematic relation between transition direction and mass ratio or orbital configuration. Mergers therefore provide one channel for perturbing the central gas and can promote either misalignment or realignment \citep{Casanueva_2022,Baker_2025a}. The outcome depends on the state of the pre-existing reservoir and subsequent gas supply, which determine whether newly supplied gas replaces or mixes with retained material \citep{Cenci_2023,Peirani_2025}.

A single-epoch misalignment census cannot distinguish recent rapid refuelling from longer-lived decoupling \citep[see also][]{Baker_2025a}, and a small present-day $\psi$ does not imply a dynamically quiet past. The nine galaxies excluded from the paired abrupt/not-abrupt comparison illustrate this point: they never enter the fiducial abrupt regime, yet some still experience extended non-aligned behaviour. Relative to the galaxies that do enter the abrupt regime, they are quieter on average, with fewer mergers, lower maximum baryonic mass ratios, and less time spent above $30^\circ$, but they are not devoid of accretion or of non-aligned evolution. Incidence and pathway frequency should therefore be treated as complementary, not interchangeable, descriptors.

We acknowledge the following limitations. Although the main paired signal is robust to the adopted thresholds, the temporal reconstruction still depends on
the chosen aperture, the treatment of unresolved snapshot gaps, and the episode-merging window. These choices may affect the detailed episode census, but are unlikely to change the sign of the paired gas-supply contrasts. The
CIELO simulations do not include active galactic nucleus (AGN) feedback. This omission is unlikely to affect the low-mass end of the sample, but AGN-driven heating and outflows could alter the state and persistence of the central gas reservoir at the upper mass end
($\log M_\star/\mathrm{M}_\odot \sim 10$--$10.7$) \citep[e.g.,][]{Davis_2011, Peirani_2025}. Despite this omission, recent work confirms that the removal of the central reservoir can be fully accounted for by starbursts, allowing kinematic misalignments to form efficiently even in the absence of AGN feedback \citep{Cenci_2023}. The merger null model in
Appendix~\ref{app:merger_linked_boundaries} preserves the merger count and tracked time baseline of each galaxy, but it cannot fully separate direct merger-boundary coupling from the broader clustering of both events during
active assembly epochs. Finally, the six CIELO zoom regions sample different large-scale environments, but the sample size does not allow a controlled environmental analysis or a clean separation between environmental trends and region-to-region variance.

In future work, we will use the particle-tracking framework developed here to quantify the stellar imprints left by past non-aligned episodes in $z=0$ galaxies, including the formation and survival of counter-rotating stellar disc components.

\section{Conclusions}
\label{sec:conclusions}

We analysed 44 central galaxies from the CIELO simulations from $z=3.5$ to $z=0$, following the intrinsic angle $\psi$ between the SF gas and stellar angular momentum vectors. We identified aligned, misaligned, and counter-rotating episodes, which are referred to as kinematic classes, and tracked the gas particles newly incorporated into the central SF reservoir. Rapid reorientation was evaluated by comparing abrupt-transition intervals with the complementary non-abrupt intervals of the same galaxies.

Our main conclusions are:

\begin{enumerate}
    \item Abrupt changes in $\psi$ occur when newly added SF gas becomes both dynamically important and misaligned with the pre-existing stellar component. This is the strongest result of the full-sample analysis because it is based on within-galaxy paired comparisons, with each system measured against its own non-abrupt baseline. For the 35 galaxies with both abrupt and non-abrupt intervals, the median paired shifts are $+1.82$ in $M_{\rm new}/M_{\rm pre-existing\,SF}$, $+0.42$ in $M_{\rm new}/M_{\rm pre-existing\,gas}$, and $+30.5^\circ$ in $\theta_{\rm new,\star}$.

    \item High-angle additions of newly incorporated SF gas within $2\,r_{\rm opt,\star}$ are preferentially fresh-gas dominated, with a larger contribution from infall plus external transfer and a smaller contribution from local cooling. The angular separation between origin channels is
    strongest when the gas joins the central SF reservoir, rather than at first $R_{\rm vir}$ crossing. After entry, persistence and recovery are controlled mainly by replacement and retention of the central SF reservoir, not by a systematic change in the star-formation delay distribution.

    \item The present-day kinematic class is not a unique record of the earlier angular momentum history. Only 4 of the 35 galaxies aligned at $z=0$ remain aligned up to $z\sim 3.5$. 
    In total, 40 out of 44 galaxies experience at least one non-aligned episode since $z \sim 3.5$, and 31 of those later realign by $z=0$. Non-aligned episodes span a broad range of durations, with no statistically distinct duration scale relative to aligned episodes within the same galaxies.

   \item Minor and major mergers occur in excess near abrupt events at infall and first pericentre, a fact that suggests a relevant role of mergers in driving significant changes in the angular momentum content as expected. Both mass classes accompany transitions towards alignment and non-alignment, with no systematic relation between transition direction and merger mass ratio or orbital configuration. Mergers close to abrupt events enter hosts with larger angles between the stellar and SF gas angular momenta at infall, indicating stronger pre-existing misalignment.
  
\end{enumerate}

Rapid gas--stellar reorientation in this sample is therefore governed by reservoir competition. Mergers and fresh gas supply can establish or reinforce the conditions for this competition, but the final kinematic outcome is set by the mass and angular momentum of the newly added SF gas relative to the pre-existing central reservoir. Whether a galaxy becomes non-aligned, remains so, or realigns depends on whether the incoming material replaces, survives within, or mixes with the gas already in place.

\begin{acknowledgements}
JGJ, PBT, CCV, BTC acknowledge partial support from ANID BASAL project FB210003.
CCV acknowledges financial support from the ESO--Chile 2024 Joint Committee ``Surviving and disrupted dwarf galaxies''. PBT acknowledges partial funding by FONDECYT-ANID 1240465/2024. NP acknowledges support from PIP Raices 29320230100004CO and PICT Raices 2023-0002. BTC gratefully acknowledges funding by ANID (Beca Doctorado Nacional, Folio 21232155). SP acknowledges partial support from ANPCyT through grant PICT 2020-00582 and CONICET through PIP 2022-0214.  
This project has received funding from the European Union Horizon 2020 Research and Innovation Programme under Marie Sk{\l}odowska-Curie Actions (MSCA) grant agreement No. 101086388-LACEGAL. This project used the Ladgerda Cluster (FONDECYT 1200703/2020 hosted at the Institute for Astrophysics, Chile), Marenostrum 4 (Barcelona SuperComputer Centre, Spain), the NLHPC (Centro de Modelamiento Matemático, Chile), and  Geryon clusters (Centre for Astrophysics, CATA, Chile).
CCV acknowledges the use of an AI-based language assistant (Claude, Anthropic) for text editing and organizational suggestions during manuscript preparation. The scientific content and conclusions remain the authors' own.
\end{acknowledgements}

\bibliographystyle{aa}
\bibliography{biblio}

@ARTICLE{Tissera_2012,
       author = {{Tissera}, Patricia B. and {White}, Simon D.~M. and {Scannapieco}, Cecilia},
        title = "{Chemical signatures of formation processes in the stellar populations of simulated galaxies}",
      journal = {\mnras},
         year = 2012,
        month = feb,
       volume = {420},
       number = {1},
        pages = {255-270},
          doi = {10.1111/j.1365-2966.2011.20028.x},
archivePrefix = {arXiv},
       eprint = {1110.5864},
 primaryClass = {astro-ph.CO},
       adsurl = {https://ui.adsabs.harvard.edu/abs/2012MNRAS.420..255T}
}

@article{Ulrich_1975,
	doi = {10.1086/129881},
	year = 1975,
	month = {dec},
	publisher = {{IOP} Publishing},
	volume = {87},
	pages = {965},
	author = {Marie-Helene Ulrich},
	title = {Velocity Gradient of the Gas Along the Apparent Minor Axis of the Galaxy {NGC} 2685},
	journal = {\pasp},}

@ARTICLE{Rubin_1992,
       author = {{Rubin}, Vera C. and {Graham}, J.~A. and {Kenney}, Jeffrey D.~P.},
        title = "{Cospatial Counterrotating Stellar Disks in the Virgo E7/S0 Galaxy NGC 4550}",
      journal = {\apjl},
         year = 1992,
        month = jul,
       volume = {394},
        pages = {L9},
          doi = {10.1086/186460},
       adsurl = {https://ui.adsabs.harvard.edu/abs/1992ApJ...394L...9R}
}

@article{Kannappan_2001,
	doi = {10.1086/318027},
	year = 2001,
	month = {jan},
	publisher = {American Astronomical Society},
	volume = {121},
	number = {1},
	pages = {140--147},
	author = {S. J. Kannappan and D. G. Fabricant},
	title = {A Broad Search for Counterrotating Gas and Stars: Evidence for Mergers and Accretion},
	journal = {\aj},}

@article{Davis_2011,
    author = {Davis, Timothy A. and Alatalo, Katherine and Sarzi, Marc and Bureau, Martin and Young, Lisa M. and Blitz, Leo and Serra, Paolo and Crocker, Alison F. and Krajnović, Davor and McDermid, Richard M. and Bois, Maxime and Bournaud, Frédéric and Cappellari, Michele and Davies, Roger L. and Duc, Pierre-Alain and de Zeeuw, P. Tim and Emsellem, Eric and Khochfar, Sadegh and Kuntschner, Harald and Lablanche, Pierre-Yves and Morganti, Raffaella and Naab, Thorsten and Oosterloo, Tom and Scott, Nicholas and Weijmans, Anne-Marie},
    title = "{The ATLAS3D project – X. On the origin of the molecular and ionized gas in early-type galaxies}",
    journal = {\mnras},
    volume = {417},
    number = {2},
    pages = {882-899},
    year = {2011},
    month = {10},issn = {0035-8711},
    doi = {10.1111/j.1365-2966.2011.19355.x},
}

@article{Barrera-Ballesteros_2014,
	author = {{Barrera-Ballesteros}, J.~K. and {Falc\'on-Barroso}, J. and {Garc\'{\i}a-Lorenzo}, B. and {van de Ven}, G. and {Aguerri}, J.~A.~L. and {Mendez-Abreu}, J. and {Spekkens}, K. and {Lyubenova}, M. and {S\'anchez}, S.~F. and {Husemann}, B. and {Mast}, D. and {Garc\'{\i}a-Benito}, R. and {Iglesias-Paramo}, J. and {del Olmo}, A. and {M\'arquez}, I. and {Masegosa}, J. and {Kehrig}, C. and {Marino}, R.~A. and {Verdes-Montenegro}, L. and {Ziegler}, B. and {McIntosh}, D.~H. and {Bland-Hawthorn}, J. and {Walcher}, C.~J.},
	title = {Kinematic alignment of non-interacting CALIFA galaxies - Quantifying the impact of bars on stellar and ionised gas velocity field orientations},
	DOI= "10.1051/0004-6361/201423488",
	url= "https://doi.org/10.1051/0004-6361/201423488",
	journal = {\aap},
	year = 2014,
	volume = 568,
	pages = "A70",
	month = "",
}

@ARTICLE{Bertola_1992,
       author = {{Bertola}, F. and {Buson}, L.~M. and {Zeilinger}, W.~W.},
        title = "{The External Origin of the Gas in S0 Galaxies}",
      journal = {\apjl},
         year = 1992,
        month = dec,
       volume = {401},
        pages = {L79},
          doi = {10.1086/186675},
       adsurl = {https://ui.adsabs.harvard.edu/abs/1992ApJ...401L..79B}
}

@article{Scannapieco_2005,
    author = {Scannapieco, C. and Tissera, P. B. and White, S. D. M. and Springel, V.},
    title = "{Feedback and metal enrichment in cosmological smoothed particle hydrodynamics simulations — I. A model for chemical enrichment}",
    journal = {\mnras},
    volume = {364},
    number = {2},
    pages = {552-564},
    year = {2005},
    month = {12},issn = {0035-8711},
    doi = {10.1111/j.1365-2966.2005.09574.x},
}

@article{Scannapieco_2006,
    author = {Scannapieco, C. and Tissera, P. B. and White, S. D. M. and Springel, V.},
    title = "{Feedback and metal enrichment in cosmological SPH simulations – II. A multiphase model with supernova energy feedback}",
    journal = {\mnras},
    volume = {371},
    number = {3},
    pages = {1125-1139},
    year = {2006},
    month = {08},issn = {0035-8711},
    doi = {10.1111/j.1365-2966.2006.10785.x},
}

@ARTICLE{Krajnovic_2006,
       author = {{Krajnovi{\'c}}, Davor and {Cappellari}, Michele and {de Zeeuw}, P. Tim and
         {Copin}, Yannick},
        title = "{Kinemetry: a generalization of photometry to the higher moments of the line-of-sight velocity distribution}",
      journal = {\mnras},
         year = "2006",
        month = "Mar",
       volume = {366},
       number = {3},
        pages = {787-802},
          doi = {10.1111/j.1365-2966.2005.09902.x},
archivePrefix = {arXiv},
       eprint = {astro-ph/0512200},
 primaryClass = {astro-ph},
       adsurl = {https://ui.adsabs.harvard.edu/abs/2006MNRAS.366..787K}
}

@ARTICLE{Pedrosa-Tissera_2015,
       author = {{Pedrosa}, Susana E. and {Tissera}, Patricia B.},
        title = "{Angular momentum evolution for galaxies in a {\ensuremath{\Lambda}}-CDM scenario}",
      journal = {\aap},
         year = 2015,
        month = dec,
       volume = {584},
          eid = {A43},
        pages = {A43},
          doi = {10.1051/0004-6361/201526440},
archivePrefix = {arXiv},
       eprint = {1508.07220},
 primaryClass = {astro-ph.GA},
       adsurl = {https://ui.adsabs.harvard.edu/abs/2015A&A...584A..43P}
}

@ARTICLE{Teklu_2015,
       author = {{Teklu}, Adelheid F. and {Remus}, Rhea-Silvia and {Dolag}, Klaus and {Beck}, Alexander M. and {Burkert}, Andreas and {Schmidt}, Andreas S. and {Schulze}, Felix and {Steinborn}, Lisa K.},
        title = "{Connecting Angular Momentum and Galactic Dynamics: The Complex Interplay between Spin, Mass, and Morphology}",
      journal = {\apj},
         year = 2015,
        month = oct,
       volume = {812},
       number = {1},
          eid = {29},
        pages = {29},
          doi = {10.1088/0004-637X/812/1/29},
archivePrefix = {arXiv},
       eprint = {1503.03501},
 primaryClass = {astro-ph.GA},
       adsurl = {https://ui.adsabs.harvard.edu/abs/2015ApJ...812...29T}
}

@ARTICLE{Lagos_2015,
       author = {{Lagos}, Claudia del P. and {Padilla}, Nelson D. and
         {Davis}, Timothy A. and {Lacey}, Cedric G. and {Baugh}, Carlton M. and
         {Gonzalez-Perez}, Violeta and {Zwaan}, Martin A. and
         {Contreras}, Sergio},
        title = "{The origin of the atomic and molecular gas contents of early-type galaxies - II. Misaligned gas accretion}",
      journal = {\mnras},
         year = "2015",
        month = "Apr",
       volume = {448},
       number = {2},
        pages = {1271-1287},
          doi = {10.1093/mnras/stu2763},
archivePrefix = {arXiv},
       eprint = {1410.5437},
 primaryClass = {astro-ph.GA},
       adsurl = {https://ui.adsabs.harvard.edu/abs/2015MNRAS.448.1271L}
}

@article{Starkenburg_2019,
   title={On the Origin of Star–Gas Counterrotation in Low-mass Galaxies},
   volume={878},
   ISSN={1538-4357},
   url={http://dx.doi.org/10.3847/1538-4357/ab2128},
   DOI={10.3847/1538-4357/ab2128},
   number={2},
   journal={\apj},
   publisher={American Astronomical Society},
   author={Starkenburg, Tjitske K. and Sales, Laura. V. and Genel, Shy and Manzano-King, Christina and Canalizo, Gabriela and Hernquist, Lars},
   year={2019},
   month={Jun},
   pages={143}
}

@ARTICLE{Khim_2020a,
       author = {{Khim}, Donghyeon J. and {Yi}, Sukyoung K. and {Pichon}, Christophe and {Dubois}, Yohan and {Devriendt}, Julien and {Choi}, Hoseung and {Bryant}, Julia J. and {Croom}, Scott M.},
        title = "{Star-Gas Misalignment in Galaxies. II. Origins Found from the Horizon-AGN Simulation}",
      journal = {\apjs},
         year = 2021,
       volume = {254},
       number = {2},
        pages = {27},
          doi = {10.3847/1538-4365/abf043}
}

@ARTICLE{Benson_2005,
       author = {{Benson}, Andrew J.},
        title = "{Orbital parameters of infalling dark matter substructures}",
      journal = {\mnras},
         year = 2005,
        month = apr,
       volume = {358},
       number = {2},
        pages = {551-562},
          doi = {10.1111/j.1365-2966.2005.08788.x},
       adsurl = {https://ui.adsabs.harvard.edu/abs/2005MNRAS.358..551B}
}

@ARTICLE{Wetzel_2011,
       author = {{Wetzel}, Andrew R.},
        title = "{On the orbits of infalling satellite haloes}",
      journal = {\mnras},
         year = 2011,
        month = mar,
       volume = {412},
       number = {1},
        pages = {49-58},
          doi = {10.1111/j.1365-2966.2010.17877.x},
       adsurl = {https://ui.adsabs.harvard.edu/abs/2011MNRAS.412...49W}
}

@ARTICLE{Jiang_2015,
       author = {{Jiang}, Lilian and {Cole}, Shaun and {Sawala}, Till and {Frenk}, Carlos S.},
        title = "{Orbital parameters of infalling satellite haloes in the hierarchical {$\\Lambda$}CDM model}",
      journal = {\mnras},
         year = 2015,
        month = apr,
       volume = {448},
       number = {2},
        pages = {1674-1686},
          doi = {10.1093/mnras/stv053},
       adsurl = {https://ui.adsabs.harvard.edu/abs/2015MNRAS.448.1674J}
}

@ARTICLE{vandeVoort_2015,
       author = {{van de Voort}, Freeke and {Davis}, Timothy A. and {Kere{\v{s}}}, Du{\v{s}}an and {Quataert}, Eliot and {Faucher-Gigu{\`e}re}, Claude-Andr{\'e} and {Hopkins}, Philip F.},
        title = "{The creation and persistence of a misaligned gas disc in a simulated early-type galaxy}",
      journal = {\mnras},
         year = 2015,
        month = aug,
       volume = {451},
       number = {3},
        pages = {3269-3277},
          doi = {10.1093/mnras/stv1217},
archivePrefix = {arXiv},
       eprint = {1504.03685},
 primaryClass = {astro-ph.GA},
       adsurl = {https://ui.adsabs.harvard.edu/abs/2015MNRAS.451.3269V}
}

@ARTICLE{Li_2021,
       author = {{Li}, Song-lin and {Shi}, Yong and {Bizyaev}, Dmitry and {Duckworth}, Christopher and {Yan}, Ren-bin and {Chen}, Yan-mei and {Bing}, Long-ji and {Chen}, Jian-hang and {Yu}, Xiao-ling and {Riffel}, Rogemar A.},
        title = "{The impact of merging on the origin of kinematically misaligned and counter-rotating galaxies in MaNGA}",
      journal = {\mnras},
         year = 2021,
        month = jan,
       volume = {501},
       number = {1},
        pages = {14-23},
          doi = {10.1093/mnras/staa3618},
archivePrefix = {arXiv},
       eprint = {1912.04522},
 primaryClass = {astro-ph.GA},
       adsurl = {https://ui.adsabs.harvard.edu/abs/2021MNRAS.501...14L}
}

@article{Casanueva_2022,
    author = {Casanueva, Catalina~I. and Lagos, Claudia~del~P. and Padilla, Nelson~D. and Davison, Thomas~A.},
    title = "{The origin of star--gas misalignments in simulated galaxies}",
    journal = {\mnras},
    year = {2022},
    volume = {514},
    number = {2},
    pages = {2031-2048},
    month = {feb},
    doi = {10.1093/mnras/stac523},
}

@article{Hahn-Abel_2011,
    author = {Hahn, Oliver and Abel, Tom},
    title = "{Multi-scale initial conditions for cosmological simulations}",
    journal = {\mnras},
    volume = {415},
    number = {3},
    pages = {2101-2121},
    year = {2011},
    month = {08},issn = {0035-8711},
    doi = {10.1111/j.1365-2966.2011.18820.x},
}

@article{Behroozi_2012,
	doi = {10.1088/0004-637x/762/2/109},
	year = 2012,
	month = {dec},
	publisher = {American Astronomical Society},
	volume = {762},
	number = {2},
	pages = {109},
	author = {Peter S. Behroozi and Risa H. Wechsler and Hao-Yi Wu},
	title = {{THE} {ROCKSTAR} {PHASE}-{SPACE} {TEMPORAL} {HALO} {FINDER} {AND} {THE} {VELOCITY} {OFFSETS} {OF} {CLUSTER} {CORES}},
	journal = {\apj},}

@article{Zhou_2022,
    author = {Zhou, Yuren and Chen, Yanmei and Shi, Yong and Bizyaev, Dmitry and Guo, Hong and Bao, Min and Xu, Haitong and Yu, Xiaoling and Brownstein, Joel R},
    title = {SDSS-IV MaNGA: global properties of kinematically misaligned galaxies},
    journal = {\mnras},
    volume = {515},
    number = {4},
    pages = {5081-5093},
    year = {2022},
    month = {07},issn = {0035-8711},
    doi = {10.1093/mnras/stac2016},
}

@article{Zhou_2024,
doi = {10.3847/1538-4357/ad8c3d},
year = {2024},
month = {dec},
publisher = {The American Astronomical Society},
volume = {977},
number = {1},
pages = {62},
author = {Zhou, Yuren and Chen, Yanmei and Shi, Yong and Gu, Qiusheng and Wang, Junfeng and Bizyaev, Dmitry},
title = {Misaligned Gas Acquisition as a Formation Pathway of S0 Galaxies},
journal = {\apj},}

@article{Zinchenko_2023,
	author = {{Zinchenko}, I.~A.},
	title = {Gas and stellar kinematic misalignment in MaNGA galaxies: What is the origin of counter-rotating gas?},
	DOI= "10.1051/0004-6361/202346846",
	url= "https://doi.org/10.1051/0004-6361/202346846",
	journal = {\aap},
	year = 2023,
	volume = 674,
	pages = "L7",
}

@article{Peirani_2025,
	author = {{Peirani}, S{\'e}bastien and {Suto}, Yasushi and {Han}, Seongbong and {Yi}, Sukyoung K. and {Dubois}, Yohan and {Kraljic}, Katarina and {Park}, Minjung and {Pichon}, Christophe},
	title = {Dissecting the formation of gas-versus-star counter-rotating galaxies from the NewHorizon simulation},
	DOI= "10.1051/0004-6361/202453577",
	url= "https://doi.org/10.1051/0004-6361/202453577",
	journal = {\aap},
	year = 2025,
	volume = 696,
	pages = "A45",
}

@article{Han_2024,
  author = {Han, Seongbong and Jang, J. K. and Contini, Emanuele and Dubois, Yohan and Jeon, Seyoung and Kaviraj, Sugata and Kimm, Taysun and Kraljic, Katarina and Oh, Sree and Peirani, S{\'e}bastien and Pichon, Christophe and Yi, Sukyoung K.},
  title = {Exploring Lenticular Galaxy Formation in Field Environments Using NewHorizon: Evidence for Counterrotating Gas Accretion as a Formation Channel},
  journal = {\apj},
  year = {2024},
  month = {dec},
  volume = {977},
  number = {1},
  pages = {116},
  doi = {10.3847/1538-4357/ad8ba7}
}

@article{Bao_2025,
  author = {Bao, Min and Chen, Yanmei and Gu, Qiusheng and Wang, Huiyuan and Shi, Yong and Wang, Peng},
  title = {The Large-scale Structure Supplies the Formation of Gas-star Misaligned Galaxies},
  journal = {\apjl},
  year = {2025},
  month = {mar},
  volume = {982},
  pages = {L29},
  doi = {10.3847/2041-8213/adbc68},
  eprint = {2503.02219},
  archivePrefix = {arXiv},
  primaryClass = {astro-ph.GA}
}

@article{Tissera_2025,
	author = {{Tissera}, Patricia B. and {Bignone}, Lucas and {Gonzalez-Jara}, Jenny and {Mu{\~n}oz-Escobar}, Ignacio and {Cataldi}, Pedro and {Miranda}, Valentina P. and {Barrientos-Acevedo}, Daniela and {Tapia-Contreras}, Brian and {Pedrosa}, Susana and {Padilla}, Nelson and {Dominguez-Tenreiro}, Rosa and {Casanueva-Villarreal}, Catalina and {Sillero}, Emanuel and {Silva-Mella}, Benjamin and {Shailesh}, Isha and {Jara-Ferreira}, Francisco},
	title = {The CIELO project: The chemo-dynamical properties of galaxies and the cosmic web},
	DOI= "10.1051/0004-6361/202453348",
	url= "https://doi.org/10.1051/0004-6361/202453348",
	journal = {\aap},
	year = 2025,
	volume = 697,
	pages = "A134",
}

@article{Fall_1980,
  author = {Fall, S.~M. and Efstathiou, G.},
  title = {Formation and rotation of disc galaxies with haloes.},
  journal = {\mnras},
  year = {1980},
  month = {oct},
  volume = {193},
  pages = {189--206},
  doi = {10.1093/mnras/193.2.189}
}

@article{Mo_1998,
  author = {Mo, H.~J. and Mao, Shude and White, Simon D.~M.},
  title = {The formation of galactic discs},
  journal = {\mnras},
  year = {1998},
  month = {apr},
  volume = {295},
  number = {2},
  pages = {319--336},
  doi = {10.1046/j.1365-8711.1998.01227.x},
  eprint = {astro-ph/9707093},
  archivePrefix = {arXiv},
  primaryClass = {astro-ph}
}

@article{Bryant_2019,
  author = {Bryant, J.~J. and Croom, S.~M. and van de Sande, J. and Scott, N. and Fogarty, L.~M.~R. and Bland-Hawthorn, J. and Bloom, J.~V. and Taylor, E.~N. and Brough, S. and Robotham, A. and Cortese, L. and Couch, W. and Owers, M.~S. and Medling, A.~M. and Federrath, C. and Bekki, K. and Richards, S.~N. and Lawrence, J.~S. and Konstantopoulos, I.~S.},
  title = {The SAMI Galaxy Survey: stellar and gas misalignments and the origin of gas in nearby galaxies},
  journal = {\mnras},
  year = {2019},
  month = {feb},
  volume = {483},
  number = {1},
  pages = {458--479},
  doi = {10.1093/mnras/sty3122},
  eprint = {1811.09298},
  archivePrefix = {arXiv},
  primaryClass = {astro-ph.GA}
}

@article{Duckworth_2020,
  author = {Duckworth, Christopher and Tojeiro, Rita and Kraljic, Katarina},
  title = {Decoupling the rotation of stars and gas -- I. The relationship with morphology and halo spin},
  journal = {\mnras},
  year = {2020},
  month = {feb},
  volume = {492},
  number = {2},
  pages = {1869--1886},
  doi = {10.1093/mnras/stz3575},
  eprint = {1910.10744},
  archivePrefix = {arXiv},
  primaryClass = {astro-ph.GA}
}

@article{Baker_2025a,
    author = {Baker, Maximilian K and Davis, Timothy A and van de Voort, Freeke and Ruffa, Ilaria},
    title = {Stellar-gas kinematic misalignments in eagle: lifetimes and longevity of misaligned galaxies},
    journal = {\mnras},
    volume = {541},
    number = {1},
    pages = {494-515},
    year = {2025},
    month = {07},
    issn = {0035-8711},
    doi = {10.1093/mnras/staf977},
    url = {https://doi.org/10.1093/mnras/staf977},
    eprint = {https://academic.oup.com/mnras/article-pdf/541/1/494/63485349/staf977.pdf},
}

@article{Ristea_2022,
  author = {Ristea, A. and Cortese, L. and Fraser-McKelvie, A. and Brough, S. and Bryant, J. J. and Catinella, B. and Croom, S. M. and Groves, B. and Richards, S. N. and van de Sande, J. and Bland-Hawthorn, J. and Owers, M. S. and Lawrence, J. S.},
  title = {The SAMI Galaxy Survey: physical drivers of stellar-gas kinematic misalignments in the nearby Universe},
  journal = {\mnras},
  year = {2022},
  month = {dec},
  volume = {517},
  number = {2},
  pages = {2677--2696},
  doi = {10.1093/mnras/stac2839}
}

@article{Cenci_2023,
  author = {Cenci, Elia and Feldmann, Robert and Gensior, Jindra and Bullock, James S. and Moreno, Jorge and Bassini, Luigi and Bernardini, Mauro},
  title = {Starburst-induced gas-star kinematic misalignment},
  journal = {\apjl},
  year = {2024},
  volume = {961},
  number = {2},
  pages = {L40},
  doi = {10.3847/2041-8213/ad1ffb}
}

@article{Persic_1996,
  author = {Persic, Massimo and Salucci, Paolo and Stel, Fulvio},
  title = {The universal rotation curve of spiral galaxies -- I. The dark matter connection},
  journal = {\mnras},
  year = {1996},
  volume = {281},
  number = {1},
  pages = {27--47},
  doi = {10.1093/mnras/281.1.27}
}

@article{Power_2003,
  author = {Power, Chris and Navarro, Julio F. and Jenkins, Adrian and Frenk, Carlos S. and White, Simon D. M. and Springel, Volker and Stadel, Joachim and Quinn, Thomas},
  title = {The inner structure of {$\Lambda$CDM} haloes -- I. A numerical convergence study},
  journal = {\mnras},
  year = {2003},
  volume = {338},
  number = {1},
  pages = {14--34},
  doi = {10.1046/j.1365-8711.2003.05925.x}
}

@misc{CasanuevaVillarreal_2026_pipeline,
  author = {Casanueva-Villarreal, C.},
  title = {CIELO Merger-Tree Cleaning Pipeline},
  year = {2026},
  howpublished = {Zenodo},
  note = {https://doi.org/10.5281/zenodo.20077625}
}

@article{TapiaContreras_2025,
  author = {Tapia-Contreras, Brian and Tissera, Patricia B. and Sillero, Emanuel and Gonzalez-Jara, Jenny and Casanueva-Villarreal, Catalina and Pedrosa, Susana and Bignone, Lucas and Padilla, Nelson D. and Dominguez-Tenreiro, Rosa},
  title = {Insight into the physical processes that shape the metallicity profiles in galaxies},
  journal = {\aap},
  year = {2025},
  volume = {700},
  pages = {A69},
  doi = {10.1051/0004-6361/202554013}
}

@article{GonzalezJara_2025_chemhalos,
  author = {Gonzalez-Jara, Jenny and Tissera, Patricia B. and Monachesi, Antonela and Sillero, Emanuel and Pallero, Diego and Pedrosa, Susana and Tau, Elisa A. and Tapia-Contreras, Brian and Bignone, Lucas},
  title = {Unveiling the formation channels of stellar halos through their chemical fingerprints},
  journal = {\aap},
  year = {2025},
  volume = {693},
  pages = {A282},
  doi = {10.1051/0004-6361/202452639}
}

@article{Knollmann_2009,
doi = {10.1088/0067-0049/182/2/608},
url = {https://doi.org/10.1088/0067-0049/182/2/608},
year = {2009},
month = {may},
publisher = {The American Astronomical Society},
volume = {182},
number = {2},
pages = {608},
author = {Knollmann, Steffen R. and Knebe, Alexander},
title = {Ahf: AMIGA'S HALO FINDER},
journal = {The Astrophysical Journal Supplement Series}
}

@ARTICLE{Springel_2001,
       author = {{Springel}, Volker and {White}, Simon D.~M. and {Tormen}, Giuseppe and {Kauffmann}, Guinevere},
        title = "{Populating a cluster of galaxies - I. Results at z=0}",
      journal = {\mnras},
         year = 2001,
        month = dec,
       volume = {328},
       number = {3},
        pages = {726-750},
          doi = {10.1046/j.1365-8711.2001.04912.x},
archivePrefix = {arXiv},
       eprint = {astro-ph/0012055},
 primaryClass = {astro-ph},
       adsurl = {https://ui.adsabs.harvard.edu/abs/2001MNRAS.328..726S}
}

@ARTICLE{Dolag_2009,
       author = {{Dolag}, K. and {Borgani}, S. and {Murante}, G. and {Springel}, V.},
        title = "{Substructures in hydrodynamical cluster simulations}",
      journal = {\mnras},
         year = 2009,
        month = oct,
       volume = {399},
       number = {2},
        pages = {497-514},
          doi = {10.1111/j.1365-2966.2009.15034.x},
archivePrefix = {arXiv},
       eprint = {0808.3401},
 primaryClass = {astro-ph},
       adsurl = {https://ui.adsabs.harvard.edu/abs/2009MNRAS.399..497D}
}

@ARTICLE{MunozEscobar_2026,
       author = {{Mu{\~n}oz-Escobar}, Ignacio and {Tissera}, Patricia B. and {Gonzalez-Jara}, Jenny and {Sillero}, Emanuel and {Miranda}, Valentina P. and {Pedrosa}, Susana and {Bignone}, Lucas},
        title = "{The mass-metallicity relation of bulges}",
      journal = {\aap},
         year = 2026,
        month = jan,
       volume = {705},
          eid = {A87},
        pages = {A87},
          doi = {10.1051/0004-6361/202557133},
archivePrefix = {arXiv},
       eprint = {2510.25642},
 primaryClass = {astro-ph.GA},
       adsurl = {https://ui.adsabs.harvard.edu/abs/2026A&A...705A..87M}
}

@ARTICLE{PlanckCollaborationXVI_2014,
       author = {{Planck Collaboration} and {Ade}, P.~A.~R. and {Aghanim}, N. and {Armitage-Caplan}, C. and {Arnaud}, M. and {Ashdown}, M. and {Atrio-Barandela}, F. and {Aumont}, J. and {Baccigalupi}, C. and {Banday}, A.~J. and {Barreiro}, R.~B. and {Bartlett}, J.~G. and {Battaner}, E. and {Benabed}, K. and {Beno{\^\i}t}, A. and {Benoit-L{\'e}vy}, A. and {Bernard}, J.-P. and {Bersanelli}, M. and {Bielewicz}, P. and {Bobin}, J. and {Bock}, J.~J. and {Bonaldi}, A. and {Bond}, J.~R. and {Borrill}, J. and {Bouchet}, F.~R. and {Bridges}, M. and {Bucher}, M. and {Burigana}, C. and {Butler}, R.~C. and {Calabrese}, E. and {Cappellini}, B. and {Cardoso}, J.-F. and {Catalano}, A. and {Challinor}, A. and {Chamballu}, A. and {Chary}, R.-R. and {Chen}, X. and {Chiang}, H.~C. and {Chiang}, L.-Y. and {Christensen}, P.~R. and {Church}, S. and {Clements}, D.~L. and {Colombi}, S. and {Colombo}, L.~P.~L. and {Couchot}, F. and {Coulais}, A. and {Crill}, B.~P. and {Curto}, A. and {Cuttaia}, F. and {Danese}, L. and {Davies}, R.~D. and {Davis}, R.~J. and {de Bernardis}, P. and {de Rosa}, A. and {de Zotti}, G. and {Delabrouille}, J. and {Delouis}, J.-M. and {D{\'e}sert}, F.-X. and {Dickinson}, C. and {Diego}, J.~M. and {Dolag}, K. and {Dole}, H. and {Donzelli}, S. and {Dor{\'e}}, O. and {Douspis}, M. and {Dunkley}, J. and {Dupac}, X. and {Efstathiou}, G. and {Elsner}, F. and {En{\ss}lin}, T.~A. and {Eriksen}, H.~K. and {Finelli}, F. and {Forni}, O. and {Frailis}, M. and {Fraisse}, A.~A. and {Franceschi}, E. and {Gaier}, T.~C. and {Galeotta}, S. and {Galli}, S. and {Ganga}, K. and {Giard}, M. and {Giardino}, G. and {Giraud-H{\'e}raud}, Y. and {Gjerl{\o}w}, E. and {Gonz{\'a}lez-Nuevo}, J. and {G{\'o}rski}, K.~M. and {Gratton}, S. and {Gregorio}, A. and {Gruppuso}, A. and {Gudmundsson}, J.~E. and {Haissinski}, J. and {Hamann}, J. and {Hansen}, F.~K. and {Hanson}, D. and {Harrison}, D. and {Henrot-Versill{\'e}}, S. and {Hern{\'a}ndez-Monteagudo}, C. and {Herranz}, D. and {Hildebrandt}, S.~R. and {Hivon}, E. and {Hobson}, M. and {Holmes}, W.~A. and {Hornstrup}, A. and {Hou}, Z. and {Hovest}, W. and {Huffenberger}, K.~M. and {Jaffe}, A.~H. and {Jaffe}, T.~R. and {Jewell}, J. and {Jones}, W.~C. and {Juvela}, M. and {Keih{\"a}nen}, E. and {Keskitalo}, R. and {Kisner}, T.~S. and {Kneissl}, R. and {Knoche}, J. and {Knox}, L. and {Kunz}, M. and {Kurki-Suonio}, H. and {Lagache}, G. and {L{\"a}hteenm{\"a}ki}, A. and {Lamarre}, J.-M. and {Lasenby}, A. and {Lattanzi}, M. and {Laureijs}, R.~J. and {Lawrence}, C.~R. and {Leach}, S. and {Leahy}, J.~P. and {Leonardi}, R. and {Le{\'o}n-Tavares}, J. and {Lesgourgues}, J. and {Lewis}, A. and {Liguori}, M. and {Lilje}, P.~B. and {Linden-V{\o}rnle}, M. and {L{\'o}pez-Caniego}, M. and {Lubin}, P.~M. and {Mac{\'\i}as-P{\'e}rez}, J.~F. and {Maffei}, B. and {Maino}, D. and {Mandolesi}, N. and {Maris}, M. and {Marshall}, D.~J. and {Martin}, P.~G. and {Mart{\'\i}nez-Gonz{\'a}lez}, E. and {Masi}, S. and {Massardi}, M. and {Matarrese}, S. and {Matthai}, F. and {Mazzotta}, P. and {Meinhold}, P.~R. and {Melchiorri}, A. and {Melin}, J.-B. and {Mendes}, L. and {Menegoni}, E. and {Mennella}, A. and {Migliaccio}, M. and {Millea}, M. and {Mitra}, S. and {Miville-Desch{\^e}nes}, M.-A. and {Moneti}, A. and {Montier}, L. and {Morgante}, G. and {Mortlock}, D. and {Moss}, A. and {Munshi}, D. and {Murphy}, J.~A. and {Naselsky}, P. and {Nati}, F. and {Natoli}, P. and {Netterfield}, C.~B. and {N{\o}rgaard-Nielsen}, H.~U. and {Noviello}, F. and {Novikov}, D. and {Novikov}, I. and {O'Dwyer}, I.~J. and {Osborne}, S. and {Oxborrow}, C.~A. and {Paci}, F. and {Pagano}, L. and {Pajot}, F. and {Paladini}, R. and {Paoletti}, D. and {Partridge}, B. and {Pasian}, F. and {Patanchon}, G. and {Pearson}, D. and {Pearson}, T.~J. and {Peiris}, H.~V. and {Perdereau}, O. and {Perotto}, L. and {Perrotta}, F. and {Pettorino}, V. and {Piacentini}, F. and {Piat}, M. and {Pierpaoli}, E. and {Pietrobon}, D. and {Plaszczynski}, S. and {Platania}, P. and {Pointecouteau}, E.},
        title = "{Planck 2013 results. XVI. Cosmological parameters}",
      journal = {\aap},
         year = 2014,
        month = nov,
       volume = {571},
          eid = {A16},
        pages = {A16},
          doi = {10.1051/0004-6361/201321591},
archivePrefix = {arXiv},
       eprint = {1303.5076},
 primaryClass = {astro-ph.CO},
       adsurl = {https://ui.adsabs.harvard.edu/abs/2014A&A...571A..16P}
}

\begin{appendix} 
\onecolumn
\section{Sample overview and angular-offset robustness}
\label{app:sample_overview}

This appendix provides the basic $z=0$ properties of the final sample and documents the stability of our kinematic classification against alternative angular-offset definitions.

\subsection{$z=0$ properties of the final sample}

Table~\ref{tab:final_sample_detail} lists the stellar, total gas, and SF gas masses within $2\,r_{\rm opt,\star}$, together with the SF gas particle count, fiducial angle, and kinematic class at $z=0$. Across the 44 galaxies, the median total gas mass is $1.54\times10^9\,\mathrm{M}_\odot$, and non-star-forming (nonSF) gas represents a median fraction of 0.38 of the central gas reservoir, with 16th and 84th percentiles of 0.13 and 0.60.

\begin{table}[H]
\caption{Properties of the final sample galaxies at $z=0$.}
\label{tab:final_sample_detail}
\centering
\small
\begin{tabular*}{\textwidth}{@{\extracolsep{\fill}}lcccccccc}
\hline\hline
Simulation & ID & Res. & $\log(M_{\star,2r_{\rm opt,\star}}/\mathrm{M}_\odot)$ & $\log(M_{{\rm gas},2r_{\rm opt,\star}}/\mathrm{M}_\odot)$ & $\log(M_{{\rm SFgas},2r_{\rm opt,\star}}/\mathrm{M}_\odot)$ & $N_{{\rm gas,SF},2r_{\rm opt,\star}}$ & $\psi$ [deg] & Class \\
\hline
P2 & 0 & L11 & 9.18 & 9.24 & 8.98 & 3413 & 7.4 & A \\
P2 & 214 & L11 & 9.86 & 9.63 & 9.49 & 10948 & 1.2 & A \\
\hline
P3 & 0 & L11 & 9.28 & 9.24 & 8.81 & 2293 & 108.9 & M \\
P3 & 232 & L11 & 10.22 & 8.51 & 8.31 & 743 & 1.0 & A \\
P3 & 271 & L11 & 9.52 & 9.07 & 8.60 & 1419 & 56.0 & M \\
P3 & 298 & L11 & 9.01 & 8.82 & 8.48 & 1125 & 149.6 & M \\
P3 & 875 & L11 & 9.46 & 9.17 & 7.75 & 199 & 1.2 & A \\
\hline
P4 & 0 & L11 & 10.12 & 8.94 & 8.82 & 2350 & 3.4 & A \\
P4 & 18 & L11 & 9.28 & 8.71 & 8.52 & 1172 & 175.9 & CR \\
P4 & 349 & L11 & 10.26 & 9.58 & 9.50 & 11370 & 5.7 & A \\
P4 & 428 & L11 & 10.09 & 9.19 & 8.97 & 3471 & 175.0 & CR \\
P4 & 1248 & L11 & 9.77 & 8.87 & 8.24 & 646 & 4.9 & A \\
P4 & 1252 & L11 & 9.65 & 9.01 & 8.71 & 1808 & 4.8 & A \\
P4 & 1258 & L11 & 9.39 & 9.44 & 9.24 & 6018 & 3.4 & A \\
\hline
P7 & 0 & L12 & 9.73 & 9.01 & 8.71 & 15309 & 1.6 & A \\
P7 & 181 & L12 & 8.92 & 9.42 & 9.02 & 29829 & 0.9 & A \\
P7 & 192 & L12 & 8.81 & 8.84 & 8.60 & 11077 & 2.0 & A \\
P7 & 200 & L12 & 8.68 & 7.79 & 7.01 & 286 & 29.4 & A \\
P7 & 2389 & L12 & 10.73 & 9.32 & 9.13 & 42645 & 15.9 & A \\
P7 & 2627 & L12 & 9.67 & 9.45 & 9.09 & 35280 & 1.6 & A \\
P7 & 2696 & L12 & 9.36 & 8.45 & 8.34 & 6253 & 8.3 & A \\
P7 & 2717 & L12 & 9.10 & 8.31 & 8.23 & 4857 & 2.3 & A \\
P7 & 2774 & L12 & 8.39 & 8.25 & 8.04 & 3132 & 2.8 & A \\
P7 & 2780 & L12 & 8.04 & 8.44 & 8.31 & 5699 & 2.8 & A \\
P7 & 7805 & L12 & 10.61 & 8.96 & 8.41 & 7830 & 50.4 & M \\
P7 & 8958 & L12 & 10.17 & 9.02 & 8.27 & 5324 & 13.1 & A \\
P7 & 9110 & L12 & 8.24 & 8.10 & 7.87 & 2111 & 33.5 & M \\
\hline
LG1 & 0 & L12 & 9.66 & 10.07 & 10.01 & 37019 & 1.0 & A \\
LG1 & 32 & L12 & 9.22 & 9.84 & 9.80 & 22468 & 1.5 & A \\
LG1 & 53 & L12 & 9.64 & 9.17 & 8.98 & 3867 & 5.0 & A \\
LG1 & 81 & L12 & 9.06 & 8.65 & 8.42 & 981 & 16.5 & A \\
LG1 & 105 & L12 & 9.26 & 9.19 & 9.08 & 4419 & 0.3 & A \\
LG1 & 107 & L12 & 9.06 & 9.55 & 9.50 & 11144 & 2.1 & A \\
LG1 & 2097 & L12 & 9.67 & 9.16 & 8.94 & 3334 & 38.2 & M \\
LG1 & 2208 & L12 & 9.40 & 9.45 & 9.41 & 9466 & 1.7 & A \\
LG1 & 2250 & L12 & 9.53 & 9.64 & 9.62 & 15795 & 3.2 & A \\
LG1 & 2298 & L12 & 9.21 & 9.44 & 9.38 & 8896 & 0.5 & A \\
LG1 & 4337 & L12 & 10.71 & 9.49 & 9.45 & 10805 & 1.3 & A \\
LG1 & 4469 & L12 & 9.91 & 9.71 & 9.63 & 16556 & 2.8 & A \\
LG1 & 4647 & L12 & 9.02 & 9.55 & 9.54 & 11954 & 2.8 & A \\
\hline
LG2 & 190 & L12 & 10.16 & 9.40 & 9.09 & 559 & 158.2 & CR \\
LG2 & 212 & L12 & 10.12 & 9.37 & 9.20 & 743 & 4.6 & A \\
LG2 & 220 & L12 & 9.97 & 9.71 & 9.49 & 1400 & 2.5 & A \\
LG2 & 238 & L12 & 9.14 & 9.19 & 8.80 & 274 & 7.3 & A \\
\hline
\end{tabular*}
\tablefoot{Columns list, from left to right, the host simulation, the galaxy identifier at $z=0$, the numerical resolution of the run, the stellar mass, the total gas mass, the SF gas mass, the number of SF gas particles, the fiducial intrinsic angular offset $\psi$, and the corresponding kinematic class. All masses and particle numbers are measured within $2\,r_{\rm opt,\star}$. Class labels are A = aligned, M = misaligned, and CR = counter-rotating. P labels denote Pehuen simulations, whereas LG labels denote Local Group analogues.}
\end{table}

Across the 2859 valid snapshots entering the time-resolved statistics, the median nonSF contribution to the sum of the gas angular momentum magnitudes is $|\mathbf{J}_{\rm nonSF}|/(|\mathbf{J}_{\rm SF}|+|\mathbf{J}_{\rm nonSF}|)=0.29$. The median angle between the SF and total gas vectors is $2.17^\circ$, and using the total gas vector changes the kinematic class in 124 snapshots (4.3\%).

Axis stability was tested using 64 complementary random particle splits per snapshot. For each snapshot, we adopted the median subset-axis separation. For the SF gas, this separation has a sample median of $0.4^\circ$ and a maximum of $23.3^\circ$. Nine stellar measurements (0.3\%) exceed $30^\circ$; treating them as unresolved leaves the episode census unchanged and preserves the signs and significance of the main paired contrasts.

\subsection{Robustness to alternative angular-offset definitions}

\begin{table}[H]
\caption{Comparison of the fiducial intrinsic angular-offset definition with the intrinsic disc-stars and all projected 2D alternatives for the final sample at $z=0$.}
\label{tab:pa_defs_delta}
\centering
\small
\begin{tabular*}{\textwidth}{@{\extracolsep{\fill}}lcccccc}
\hline\hline
Simulation & ID & $\psi$ [deg] & $\Delta_{\rm 3D,disc}$ & $\Delta_{\rm 2D,xy}$ & $\Delta_{\rm 2D,xz}$ & $\Delta_{\rm 2D,yz}$ \\
\hline
P2  & 0    & 7.39   & +1.37        & -0.41        & -4.54         & -2.94 \\
P2  & 214  & 1.22   & -0.29        & +0.12        & +1.55         & -1.03 \\
\hline
P3  & 0    & 108.90 & +7.85        & -67.94       & +6.93         & +26.45 \\
P3  & 232  & 0.98   & +0.89        & +0.62        & -0.71         & +0.14 \\
P3  & 271  & 56.01  & +1.01        & -13.93       & +5.09         & -12.36 \\
P3  & 298  & 149.56 & +2.47$^\ast$ & -20.60       & -1.44         & +18.12$^\ast$ \\
P3  & 875  & 1.22   & -0.39        & +0.01        & -0.26         & -1.02 \\
\hline
P4  & 0    & 3.40   & +0.35        & -0.02        & -1.77         & -3.05 \\
P4  & 18   & 175.90 & -2.00        & +2.25        & -0.65         & +0.78 \\
P4  & 349  & 5.72   & +0.66        & -0.87        & +1.91         & -2.45 \\
P4  & 428  & 175.04 & +1.87        & +1.47        & +0.34         & +3.86 \\
P4  & 1248 & 4.87   & +2.29        & -0.19        & +9.30         & -3.31 \\
P4  & 1252 & 4.75   & -0.58        & -4.20        & -0.01         & +0.89 \\
P4  & 1258 & 3.40   & -0.27        & -1.80        & -1.78         & -0.26 \\
\hline
P7  & 0    & 1.61   & -0.08        & +1.09        & -1.48         & +0.12 \\
P7  & 181  & 0.94   & -0.76        & +0.01        & -0.87         & -0.11 \\
P7  & 192  & 1.99   & -0.19        & +0.03        & -1.98         & +9.54 \\
P7  & 200  & 29.42  & -0.37        & -26.75       & +5.47$^\ast$  & +9.60$^\ast$ \\
P7  & 2389 & 15.86  & +3.27        & +1.88        & -13.97        & +16.40$^\ast$ \\
P7  & 2627 & 1.56   & +0.23        & +0.00        & -0.80         & -0.95 \\
P7  & 2696 & 8.33   & -0.97        & -0.26        & -8.05         & -0.76 \\
P7  & 2717 & 2.34   & -0.31        & +0.47        & +1.16         & -1.53 \\
P7  & 2774 & 2.83   & +1.52        & -2.31        & +8.31         & -0.00 \\
P7  & 2780 & 2.80   & -0.41        & -1.01        & -0.51         & +3.27 \\
P7  & 7805 & 50.39  & -1.66        & -10.21       & -33.56$^\ast$ & -3.43 \\
P7  & 8958 & 13.14  & +0.57        & +8.12        & -3.95         & -3.36 \\
P7  & 9110 & 33.49  & +7.89        & -0.32        & -27.29$^\ast$ & +115.73 \\
\hline
LG1 & 0    & 1.02   & -0.68        & -0.05        & -0.89         & -0.06 \\
LG1 & 32   & 1.52   & -0.09        & -1.15        & +0.02         & +1.08 \\
LG1 & 53   & 5.02   & +0.15        & -3.07        & -0.14         & -0.88 \\
LG1 & 81   & 16.48  & -10.33       & +84.58$^\ast$& -0.14         & -14.10 \\
LG1 & 105  & 0.29   & +0.43        & -0.07        & -0.07         & +0.08 \\
LG1 & 107  & 2.07   & -0.57        & -0.55        & +2.68         & -0.64 \\
LG1 & 2097 & 38.16  & -2.56        & +9.53        & +8.77         & -31.32$^\ast$ \\
LG1 & 2208 & 1.67   & -0.50        & -1.55        & -0.04         & -1.16 \\
LG1 & 2250 & 3.22   & -0.81        & -0.01        & -0.65         & -2.96 \\
LG1 & 2298 & 0.51   & -0.09        & -0.21        & +0.11         & -0.08 \\
LG1 & 4337 & 1.31   & +0.74        & +0.04        & -0.37         & -1.07 \\
LG1 & 4469 & 2.77   & +0.57        & -0.19        & -0.79         & -0.99 \\
LG1 & 4647 & 2.78   & -0.75        & -2.56        & +2.00         & +0.55 \\
\hline
LG2 & 190  & 158.22 & +7.00        & +5.79        & +9.98         & +5.25 \\
LG2 & 212  & 4.62   & -0.57        & +0.17        & -4.09         & +7.83 \\
LG2 & 220  & 2.48   & -0.30        & +0.45        & -1.50         & -0.37 \\
LG2 & 238  & 7.32   & -0.33        & -4.61        & +10.85        & -0.27 \\
\hline
\end{tabular*}
\tablefoot{Comparison of the fiducial 3D angle ($\psi$) against alternative definitions testing stellar tracer sensitivity ($\Delta_{\rm 3D,disc}$) and viewing geometry ($\Delta_{\rm 2D,xy}$, $\Delta_{\rm 2D,xz}$, $\Delta_{\rm 2D,yz}$), as defined in Sect.~\ref{subsec:misalignment_metric}. Tabulated $\Delta$ values denote the difference between the alternative angle and $\psi$. Positive values indicate the alternative angle is larger. Asterisks ($^\ast$) flag cases where the alternative definition alters the kinematic class. As expected, class changes are rare and occur exclusively near the adopted $30^\circ$ and $150^\circ$ boundaries or in sensitive geometric projections.}
\end{table}

Table~\ref{tab:pa_defs_delta} compares the fiducial intrinsic angle, $\psi$, against both the alternative disc-based definition and the three orthogonal 2D projections ($xy$, $xz$, and $yz$). As discussed in Sect.~\ref{subsec:misalignment_metric}, the underlying classification is highly stable. The few observed class changes are strictly confined to two configurations:

\begin{itemize}
    \item Boundary cases: Galaxies lying within a few degrees of the $30^\circ$ or $150^\circ$ thresholds (e.g., P3--gx298, P7--gx200, and P7--gx2389) can cross into an adjacent class due to minor definition variations.
    \item Small projected angular momentum vectors: In the affected viewing directions, at least one of the stellar or SF gas angular momentum vectors lies close to the line-of-sight axis, producing a small projected component and making the inferred 2D angle sensitive to small transverse variations. This geometric effect produces the largest deviations, including LG1--gx81 ($+84.6^\circ$ in $xy$), P7--gx7805 ($-33.6^\circ$ in $xz$), and LG1--gx2097 ($-31.3^\circ$ in $yz$).
\end{itemize}
These anomalies are purely geometric or boundary-driven, confirming that the fiducial classification is not biased by the choice of stellar tracer and behaves exactly as expected under projection.

\section{Case-study merger context and supporting properties}
\label{app:case_merger_support}
\begin{figure}[H] 
\centering
\includegraphics[width=\textwidth]{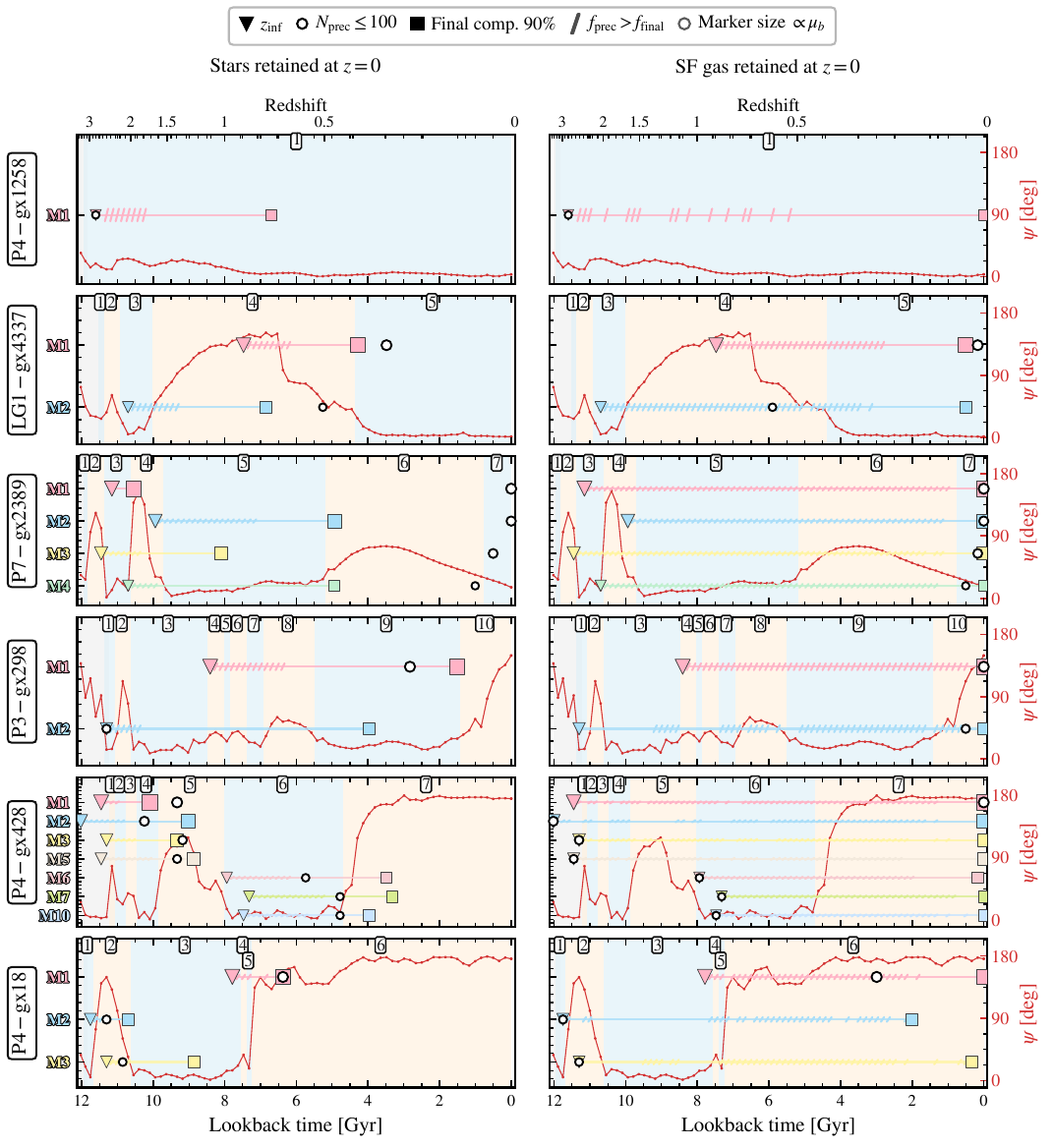}
\caption{Phase transfer for mergers contributing at least 5\% to either the merger-origin stellar or merger-origin SF gas component at $z=0$. Left and right panels show the retained stellar and SF gas components, respectively. Merger labels are assigned within each galaxy in decreasing baryonic mass ratio, $\mu_b$. Triangles mark infall, circles mark the time when the precursor component falls below 100 particles, and squares mark the time when the final component reaches 90\% of its retained $z=0$ mass. The precursor component denotes progenitor gas for the retained stellar component in the left panel and nonSF gas for the retained SF gas component in the right panel. Slanted tick marks indicate snapshots in which the precursor component still exceeds the corresponding final component. Marker size scales with $\mu_b$. The red curve shows $\psi$, and the shaded bands mark aligned (blue), non-aligned (orange), and omitted (grey) intervals. Numbered labels give the episode sequence from high to low redshift. Table~\ref{tab:case_merger_diagnostics} lists the corresponding merger timings, orbital classes, settling lags, and retained $z=0$ contributions.}
\label{fig:phase_handoff_cases}
\end{figure}

This appendix provides the event-level merger diagnostics supporting the six representative systems discussed in Sect.~\ref{sec:case_mergers}. Figure~\ref{fig:phase_handoff_cases} and Table~\ref{tab:case_merger_diagnostics} detail the transfer histories, orbital geometries, and $z=0$ retained fractions of the merger-linked stellar and SF gas components. These diagnostics show that the stellar and gaseous materials from the same merger do not settle into the central galaxy simultaneously; the stellar component systematically reaches its final retained mass earlier than the SF gas ($\Delta t_{90} > 0$). This confirms the main-text finding that gas-phase reorientation and stellar assembly are not strictly synchronous.
The orbital labels and transfer diagnostics used in Fig.~\ref{fig:phase_handoff_cases} and Table~\ref{tab:case_merger_diagnostics} are defined below.

\begin{table}[H]
\caption{Key merger diagnostics for all mergers shown in Fig.~\ref{fig:phase_handoff_cases}.}
\label{tab:case_merger_diagnostics}
\centering
\small
\setlength{\tabcolsep}{3.2pt}
\begin{tabular*}{\textwidth}{@{\extracolsep{\fill}}llccccccccc}
\hline\hline
Galaxy & Merger &
Stellar &
SF gas &
$\mu_b$ & $f_{\rm g}$ & Geometry &
$t_{\rm inf}/t_{\rm peri1}$ &
$\Delta \mathrm{PA}_{\rm peri1-inf}$ &
$\Delta t_{90}$ &
$f_{\star,z=0}/f_{{\rm SF},z=0}$ \\
& & episode(s) & episode(s) & & & & [Gyr] & [deg] & [Gyr] & [\%] \\
\hline
P4--gx1258  & M1  & Ep.~1      & Ep.~1      & 0.011 & 0.93 & polar/rad   & 11.60/11.30 &  -8.4 &  6.69 &  0.17/0.36 \\
LG1--gx4337 & M1  & Ep.~4--5   & Ep.~4--5   & 0.190 & 0.37 & polar/rad   &  7.48/ 6.85 &  +6.5 &  3.79 & 10.43/10.40 \\
LG1--gx4337 & M2  & Ep.~3--4   & Ep.~3--5   & 0.024 & 0.80 & polar/rad   & 10.70/ 9.48 & +75.5 &  6.35 &  0.65/1.05 \\
P7--gx2389  & M1  & Ep.~3--4   & Ep.~3--7   & 0.681 & 0.28 & polar/rad   & 11.15/10.85 &  +9.0 & 10.55 & 22.80/3.87 \\
P7--gx2389  & M2  & Ep.~4--6   & Ep.~4--7   & 0.132 & 0.12 & incl./mix   &  9.94/ 6.85 &  -7.0 &  4.94 & 12.13/3.96 \\
P7--gx2389  & M3  & Ep.~2--5   & Ep.~2--7   & 0.109 & 0.42 & incl./mix   & 11.45/11.15 & -89.2 &  8.10 &  2.78/0.99 \\
P7--gx2389  & M4  & Ep.~3--6   & Ep.~3--7   & 0.018 & 0.84 & polar/rad   & 10.70/ 9.18 &  -5.5 &  4.94 &  0.25/0.64 \\
P3--gx298   & M1  & Ep.~4--9   & Ep.~4--10  & 0.198 & 0.81 & CP-ret/rad  &  8.41/ 7.94 &  -9.2 &  1.50 &  5.60/5.23 \\
P3--gx298   & M2  & Ep.~1--9   & Ep.~1--10  & 0.069 & 0.97 & polar/rad   & 11.30/11.00 & +23.6 &  3.97 &  0.70/0.10 \\
P4--gx428   & M1  & Ep.~1--4   & Ep.~1--7   & 0.224 & 0.64 & CP-ret/rad  & 11.45/10.55 & +32.0 & 10.09 &  5.09/0.50 \\
P4--gx428   & M2  & Ep.~1--5   & Ep.~1--7   & 0.121 & 0.92 & CP-pro/mix  & 12.02/11.75 & -21.8 &  9.02 &  0.92/0.45 \\
P4--gx428   & M3  & Ep.~1--5   & Ep.~1--7   & 0.085 & 0.74 & CP-pro/mix  & 11.30/10.25 &  +7.1 &  9.33 &  2.54/0.34 \\
P4--gx428   & M5  & Ep.~1--5   & Ep.~1--7   & 0.056 & 0.80 & CP-pro/mix  & 11.45/11.15 & +75.1 &  8.87 &  0.88/0.24 \\
P4--gx428   & M6  & Ep.~6--7   & Ep.~6--7   & 0.024 & 0.53 & CP-ret/rad  &  7.94/ 7.32 &  -8.6 &  3.31 &  0.88/0.65 \\
P4--gx428   & M7  & Ep.~6--7   & Ep.~6--7   & 0.017 & 0.48 & CP-pro/rad  &  7.32/ 6.69 &  -0.9 &  3.32 &  0.48/0.89 \\
P4--gx428   & M10 & Ep.~6--7   & Ep.~6--7   & 0.014 & 0.86 & polar/rad   &  7.48/ 7.01 &  -0.7 &  3.97 &  0.40/0.66 \\
P4--gx18    & M1  & Ep.~3--6   & Ep.~3--6   & 0.246 & 0.73 & CP-ret/rad  &  7.79/ 6.53 & -50.0 &  6.38 &  7.38/11.49 \\
P4--gx18    & M2  & Ep.~1--2   & Ep.~1--6   & 0.061 & 0.75 & polar/rad   & 11.75/11.45 & -44.6 &  8.70 &  1.75/0.09 \\
P4--gx18    & M3  & Ep.~2--3   & Ep.~2--6   & 0.044 & 0.91 & incl./rad   & 11.30/10.85 & -88.5 &  8.54 &  0.91/0.05 \\
\hline
\end{tabular*}
\tablefoot{The table lists all mergers shown in Fig.~\ref{fig:phase_handoff_cases}. The columns ``Stellar episode(s)'' and ``SF gas episode(s)'' give the episode intervals spanned between merger infall and $t_{\star,90}$ or $t_{{\rm SF},90}$, respectively, where episode~1 is the highest-redshift episode in the plotted range and the numbering increases toward lower redshift. Geometry combines the orbital-plane class and the radial or mixed approach class from the geometry analysis. Abbreviations are CP-ret = coplanar retrograde, CP-pro = coplanar prograde, incl. = inclined, polar = polar, rad = radial-dominated, and mix = mixed. The quantity $\Delta \mathrm{PA}_{\rm peri1-inf}$ gives the change in the merger projected orientation between infall and first pericentric passage. Positive values of $\Delta t_{90}$ indicate that the stellar component linked to the merger reaches 90\% of its retained $z=0$ mass earlier than the corresponding SF gas component. The final column gives the percentages contributed by that merger to the total stellar and SF gas masses at $z=0$.}
\end{table}

\paragraph{First pericentric passage:}
We identify the first pericentric passage, $t_{\rm peri1}$, as the first post-infall snapshot at which the radial velocity changes from negative to non-negative. This criterion applies to 343 of the 395 mergers in the final catalogue and to 18 of the 19 mergers listed in Table~\ref{tab:case_merger_diagnostics}. When no such change is recovered, we adopt the first local minimum in $r_{\rm rel}$ (three cases), or the global minimum when no local minimum is identified (49 cases). The remaining merger in Table~\ref{tab:case_merger_diagnostics} is assigned from its first local minimum.

\paragraph{Compact orbital labels:}
  Table~\ref{tab:case_merger_diagnostics} summarises the orbital geometry at infall using the inclination $i_{\rm orb}\equiv\cos^{-1}(\hat{\mathbf{L}}_{\rm orb}\cdot\hat{\mathbf{J}}_{\star,\mathrm{cen}})$ and the radiality parameter $\eta_r\equiv |v_r|/v_{\rm rel}$, where $\hat{\mathbf{J}}_{\star,\mathrm{cen}}$ is the stellar
  rotation axis of the host.
  We classify orbits as coplanar prograde ($i_{\rm orb}\leq30^\circ$), coplanar retrograde ($i_{\rm orb}\geq150^\circ$), polar ($60^\circ\leq i_{\rm orb}\leq120^\circ$), and inclined otherwise. The approach is classified as radial-dominated for $\eta_r\geq0.7$, tangential-dominated for $\eta_r\leq0.3$, and mixed otherwise.

Figure~\ref{fig:phase_handoff_cases} is a timing diagram of the transfer of merger-associated material. For the retained stellar and SF gas components shown in the left and right panels, the precursors are progenitor gas and nonSF gas, respectively. Triangles mark infall, circles mark when the precursor falls below the adopted 100-particle threshold, and squares mark when the retained $z=0$ component reaches 90\% of its final mass. The relative timing of these markers compares precursor depletion with the assembly of the retained component, while the red curve traces the host $\psi$ history.

\section{Temporal-history summary}
\label{app:temporal_summary}
\begin{figure}[H] 
\centering
\includegraphics[width=\textwidth]{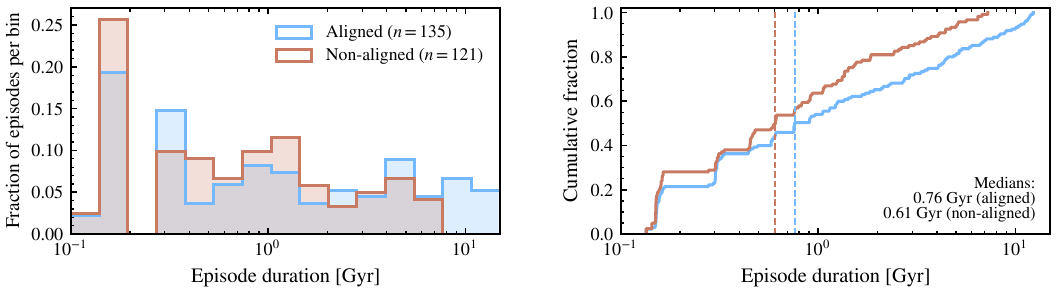}
\caption{Episode-duration distributions for the 44 galaxies. The left panel shows aligned and non-aligned episodes in common logarithmic duration bins; each histogram is normalised by the total number of episodes in its class. The right panel shows the cumulative distributions, with dashed vertical lines marking the medians.}
\label{fig:episode_duration_distributions}
\end{figure}

This appendix provides the full temporal classification of the 44-galaxy sample, expanding upon the finding that present-day kinematic classes do not uniquely reflect the preceding history of a galaxy. Table~\ref{tab:history_class_summary} categorises the sample into four broad evolutionary history classes based on their refined episode sequences. Many galaxies aligned at $z=0$ underwent substantial non-aligned periods in the past. The merger counts overlap among the four history classes. The median maximum baryonic mass ratio is $\mu_{b,\max}=0.046$ for the always-aligned class, compared with 0.319, 0.324, and 0.246 for the temporary-non-alignment, final-misaligned, and final-counter-rotating classes, respectively. These values are descriptive because the last two classes contain only six and three galaxies.
Figure~\ref{fig:episode_duration_distributions} shows the episode duration histograms and cumulative distributions. Most non-aligned episodes are short-lived, with typical durations below $1\mathrm{\,Gyr}$. Their cumulative distribution truncates earlier than that of aligned episodes, showing that non-aligned states do not reach the extreme, multi-Gyr maximum durations observed in their aligned counterparts. Despite this difference in the extreme tail, our within-galaxy paired tests confirm that the bulk of the distributions are not statistically distinct. Long-lived misalignments are therefore best interpreted as the extended tail of a continuous distribution of ultimately transient states, not as permanent kinematic features.

\begin{table}[H]
\caption{Summary of the four temporal-history classes in the 44-galaxy sample.}
\label{tab:history_class_summary}
\centering
\small
\begin{tabular}{lcccccc}
\hline\hline
History class & Count & Median $N_{\rm ep}$ & Median $f_{\rm NA}$ & Median $N_{\rm mergers}$ & Median $\mu_{b,\max}$ \\
\hline
Always aligned & 4 & 1.0 & 0.00 & 5.5 & 0.046 \\
Temporary non-alignment, finally aligned & 31 & 6.0 & 0.23 & 8.0 & 0.319 \\
Final misaligned & 6 & 7.5 & 0.62 & 3.0 & 0.324 \\
Final counter-rotating & 3 & 7.0 & 0.65 & 17.0 & 0.246 \\
\hline
\end{tabular}
\tablefoot{\raggedright The history classes are defined from the refined episode sequences. Here $f_{\rm NA}$ is the fraction of the tracked interval spent in a non-aligned state. Merger counts include the full catalogue with $\mu_{\rm b}\geq10^{-3}$, and $\mu_{b,\max}$ is the maximum baryonic merger mass ratio of each galaxy over the tracked interval. The medians are descriptive; the final-misaligned and final-counter-rotating classes contain only six and three galaxies, respectively.}
\end{table}

\section{Statistical extensions of the paired diagnostics}
\label{app:stat_extensions}
This appendix presents supporting tests for the main paired diagnostics. Section~\ref{app:operational_sensitivity} tests their stability against the adopted thresholds. Section~\ref{app:paired_subset_eligibility} documents the samples entering each comparison, while Sect.~\ref{app:counterexamples} examines departures from the dominant trends. Section~\ref{app:origin_redshift} measures the redshift dependence of the immediate-entry signal. Section~\ref{app:merger_abrupt_distance} describes the temporal separation between merger infall and abrupt events, while Sect.~\ref{app:merger_linked_boundaries} tests the proximity of merger milestones to episode boundaries and abrupt events against a galaxy-preserving null model.

\subsection{Operational sensitivity of the paired abrupt signal}
\label{app:operational_sensitivity}
Table~\ref{tab:operational_sensitivity} tests the sensitivity of the main abrupt/not-abrupt diagnostic to the two explicit threshold choices: the aligned/non-aligned boundary ($\psi_{\rm align}$) and the abrupt-event cut ($|\Delta\psi|_{\rm abrupt}$). Across the explored grid, the signs of all three gas-supply diagnostics remain unchanged and statistically significant. This confirms that the paired results reported in Sect.~\ref{subsec:results_stats} are highly robust and not an artefact of fine-tuned thresholding. 
In this diagnostic, abrupt-associated snapshots are tagged directly from $|\Delta\psi|$; therefore, changing $\psi_{\rm align}$ alters the episode census, but not the event tagging used in the paired contrast. Consequently, the meaningful variation in the table lies exclusively across the rows with different $|\Delta\psi|_{\rm abrupt}$ values.

\begin{table}[!ht]
\caption{Sensitivity of the main paired abrupt/not-abrupt signal to the alignment threshold and abrupt-event cut.}
\label{tab:operational_sensitivity}
\centering
\small
\setlength{\tabcolsep}{5pt}
\renewcommand{\arraystretch}{1.05}
\begin{tabular}{ccccccccc}
\hline\hline
$\psi_{\rm align}$ [$^\circ$] & $|\Delta\psi|_{\rm abrupt}$ [$^\circ$] & $N_{\rm gal}$ & $\Delta(M_{\rm new}/M_{\rm pre-existing\,SF})$ & $p_{\rm sign}$ & $\Delta(M_{\rm new}/M_{\rm pre-existing\,gas})$ & $p_{\rm sign}$ & $\Delta\theta_{\rm new,\star}$ [$^\circ$] & $p_{\rm sign}$ \\
\hline
20 & 25 & 37 & +1.55 & $5.53\times10^{-10}$ & +0.47 & $7.43\times10^{-6}$ & +35.3 & $4.13\times10^{-5}$ \\
20 & 35 & 35 & +1.82 & $4.18\times10^{-7}$ & +0.42 & $3.47\times10^{-6}$ & +30.5 & $3.67\times10^{-8}$ \\
20 & 45 & 30 & +1.98 & $5.95\times10^{-5}$ & +0.47 & $1.43\times10^{-3}$ & +41.1 & $8.43\times10^{-6}$ \\
30 & 25 & 37 & +1.55 & $5.53\times10^{-10}$ & +0.47 & $7.43\times10^{-6}$ & +35.3 & $4.13\times10^{-5}$ \\
30 & 35 & 35 & +1.82 & $4.18\times10^{-7}$ & +0.42 & $3.47\times10^{-6}$ & +30.5 & $3.67\times10^{-8}$ \\
30 & 45 & 30 & +1.98 & $5.95\times10^{-5}$ & +0.47 & $1.43\times10^{-3}$ & +41.1 & $8.43\times10^{-6}$ \\
40 & 25 & 37 & +1.55 & $5.53\times10^{-10}$ & +0.47 & $7.43\times10^{-6}$ & +35.3 & $4.13\times10^{-5}$ \\
40 & 35 & 35 & +1.82 & $4.18\times10^{-7}$ & +0.42 & $3.47\times10^{-6}$ & +30.5 & $3.67\times10^{-8}$ \\
40 & 45 & 30 & +1.98 & $5.95\times10^{-5}$ & +0.47 & $1.43\times10^{-3}$ & +41.1 & $8.43\times10^{-6}$ \\
\hline
\end{tabular}
\tablefoot{All entries report galaxy-level paired contrasts defined as abrupt $-$ not abrupt, with two-sided sign tests on within-galaxy medians. Across the explored grid, the signs of all three diagnostics are unchanged and remain statistically significant. These tests therefore support robustness with respect to the alignment threshold and abrupt cut. Repeated values across different $\psi_{\rm align}$ rows are expected for this specific diagnostic because abrupt tagging is driven by $|\Delta\psi|$ and the same retained episode-history mask used in the time-history map, not by class labels. The sensitivity grid is evaluated on the same 44 reconstructed $\psi$ histories and the same snapshot-level paired diagnostics used in the main figure. The fiducial row, $\psi_{\rm align}=30^\circ$ and $|\Delta\psi|_{\rm abrupt}=35^\circ$, reproduces the main-text paired shifts reported in Sect.~\ref{subsec:results_stats}.}
\end{table}

\subsection{Eligibility of the paired sub-samples}
\label{app:paired_subset_eligibility}
Table~\ref{tab:paired_subset_eligibility} summarises the eligibility of the paired sub-samples used in the main-text figures. While the episode reconstruction and the global census statistics involve the full 44-galaxy sample, the paired figures isolate smaller subsets because the specific comparisons require it. Specifically, the abrupt/not-abrupt figure selects only the galaxies that enter both temporal regimes. The fresh-fraction figure applies an additional restriction because valid immediate-entry measurements are not available for every galaxy in both angle bins, so its omissions should not be read as a calm-galaxy subset in the same sense.

\begin{table}[!ht]
\caption{Eligibility summary for the paired sub-samples used in the main-text figures.}
\label{tab:paired_subset_eligibility}
\centering
\small
\setlength{\tabcolsep}{3pt}
\renewcommand{\arraystretch}{1.05}
\begin{tabular}{>{\raggedright\arraybackslash}p{2.8cm}ccccc>{\raggedright\arraybackslash}p{4.9cm}}
\hline\hline
Sub-sample & $N_{\rm gal}$ & $z=0$ A/M/CR & Median $N_{\rm mergers}$ & Median $\mu_{b,\max}$ & Median $f_{\rm NA}$ & Selection meaning \\
\hline
Full reconstructed sample & 44 & 35/6/3 & 7.5 & 0.297 & 0.250 & All galaxies with refined episode histories \\
Main paired abrupt subset & 35 & 26/6/3 & 8.0 & 0.319 & 0.344 & Contains both abrupt-associated and not-abrupt snapshots \\
Excluded from abrupt subset & 9 & 9/0/0 & 3.0 & 0.070 & 0.011 & No abrupt-associated snapshots under the fiducial definition \\
Fresh-fraction paired subset & 24 & 20/3/1 & 8.0 & 0.358 & 0.250 & Usable origin measurements in both angle bins \\
Excluded from fresh-fraction subset & 20 & 15/3/2 & 5.5 & 0.194 & 0.269 & Missing origin measurements, or present in only one angle bin \\
\hline
\end{tabular}
\tablefoot{The third column gives the present-day class mix in the order aligned/misaligned/counter-rotating. Here $f_{\rm NA}$ denotes the fraction of the tracked interval spent in a non-aligned state. The exclusions in the two main-text figures have different meanings. For Fig.~\ref{fig:paired_signal}, the nine excluded galaxies form a physically quieter subset: all are aligned at $z=0$, none has abrupt-associated snapshots, and their typical merger counts and merger ratios are lower. For the fresh-fraction panel of Fig.~\ref{fig:support_signal}, the exclusion is different: six galaxies have no usable immediate-entry channel classification here, while the rest contribute valid measurements in only one of the two angle bins.}
\end{table}

\subsection{Counterexamples to the paired trends}
\label{app:counterexamples}
The paired signals in the main text are strong, but the systems that do not follow the dominant trends are physically informative. In Fig.~\ref{fig:paired_signal}, nine of the 35 paired galaxies show a non-positive shift in at least one diagnostic, although only five depart clearly from the overall pattern. Three galaxies (P7--gx181, P7--gx2627, and P7--gx8958) differ from the dominant trend only in $M_{\rm new}/M_{\rm pre-existing\,SF}$, while $M_{\rm new}/M_{\rm pre-existing\,gas}$ and $\theta_{\rm new,\star}$ still increase. These are denominator-driven cases in which the pre-existing SF reservoir also grows, reducing the mass contrast without removing the signature of dynamically important, tilted gas. Two others (P4--gx18 and LG2--gx190) show increased gas impact with little or no increase in $\theta_{\rm new,\star}$, consistent with dynamically important additions that remain sufficiently aligned not to amplify the stellar-offset signal.
The support diagnostics show analogous failure modes. In the upper panel of Fig.~\ref{fig:support_signal}, six of the 24 eligible galaxies have $f_{\rm fresh}(\mathrm{high\ angle}) \leq f_{\rm fresh}(\mathrm{low\ angle})$; the strongest reversals (e.g., P3--gx271) are simply more cooling-dominated in their high-angle state. Tilted gas supply is therefore not sufficient on its own: the outcome also depends on when the gas arrives and how strongly it couples to the pre-existing reservoir.

\subsection{Redshift dependence of the immediate-entry signal}
\label{app:origin_redshift}
\begin{figure}[H]
\centering
\includegraphics[width=\textwidth]{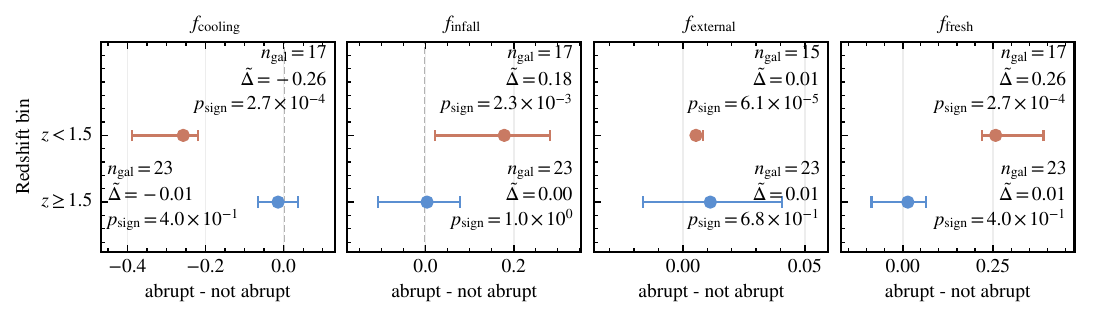}
\caption{Redshift-split extension of the abrupt/not-abrupt comparison for the immediate entry channels of the newly added SF gas. From left to right, the panels give the paired median shifts in the cooling, infall, external-transfer, and total fresh fractions, always defined as abrupt $-$ not abrupt. Each panel shows the two redshift bins, $z<1.5$ and $z\geq1.5$, as separate horizontal comparisons. Points mark the median paired shift, horizontal bars show the $95\%$ bootstrap confidence interval, and the dashed vertical line marks zero. The in-panel annotations give the number of galaxies in the paired sample and the two-sided sign-test probability.}
\label{fig:origin_redshift_split}
\end{figure}

Figure~\ref{fig:origin_redshift_split} shows the redshift-split extension of the immediate-entry signal, demonstrating that the abrupt/not-abrupt contrast is strongly redshift dependent. At $z<1.5$, abrupt-associated snapshots are substantially less cooling-dominated and more fresh-dominated than the not-abrupt snapshots of the same galaxies (e.g., $\Delta f_{\rm fresh}=+0.257$, $p=2.75\times10^{-4}$). By contrast, the same immediate-entry contrasts are consistent with zero at $z\geq1.5$.
The main-text paired signal is therefore strongest at low and intermediate redshifts, where abrupt changes preferentially coincide with a shift away from cooling-dominated supply. At higher redshift, the contrast weakens because the environment is already saturated with fresh gas, regardless of whether a snapshot lies near an abrupt change. This analysis serves as a redshift-stratified refinement of the main text, showing that the fresh-gas trigger is primarily a late-time phenomenon.

\subsection{Temporal separation between merger infall and abrupt events}
\label{app:merger_abrupt_distance}

\begin{figure}[H]
\centering
\includegraphics[width=\textwidth]{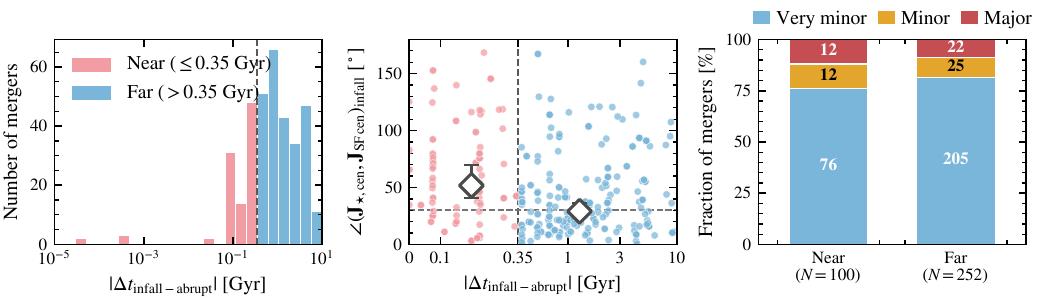}
\caption{Temporal separation between merger infall and the nearest abrupt event for the 352 mergers with a finite separation. The left panel shows the separation distribution, with salmon and blue denoting mergers within and beyond 0.35~Gyr, respectively. The middle panel shows the angle between the host stellar and SF gas angular momentum vectors at infall for the 346 mergers with measurable angles; diamonds and bars mark the medians and their 95\% bootstrap confidence intervals, and the horizontal dashed line marks $30^\circ$. The left panel uses logarithmic bins and a logarithmic separation axis. The middle panel is linear below 0.35~Gyr and logarithmic at larger separations. The right panel gives the mass-class composition of the near and far subsets, normalised separately within each subset; numbers within the bars give the absolute merger counts. Very minor, minor, and major mergers correspond to $10^{-3}\leq\mu_{\rm b}<0.1$, $0.1\leq\mu_{\rm b}<0.25$, and $\mu_{\rm b}\geq0.25$, respectively. Vertical lines mark the fiducial 0.35~Gyr interval in the left and middle panels.}
\label{fig:merger_abrupt_distance}
\end{figure}

Figure~\ref{fig:merger_abrupt_distance} shows the temporal separation, $|\Delta t_{\rm infall-abrupt}|$, between merger infall and the nearest abrupt event for the full cleaned merger catalogue. A finite separation is available for 352 of the 395 mergers: 100 lie within the fiducial 0.35~Gyr interval and 252 lie at larger separations. The remaining 43 mergers occur in nine galaxies without an identified abrupt event. The finite separations have a median of 0.73~Gyr and a mean of 1.58~Gyr. Separations shorter than the typical snapshot spacing of $\simeq0.17$~Gyr arise when an abrupt event, represented by the midpoint of two adjacent snapshots, nearly coincides with the infall snapshot; they do not imply finer temporal resolution.

The host stellar--SF gas angle is measurable for 346 mergers with finite separations. Its median is $51.8^\circ$ for the 99 mergers within 0.35~Gyr and $29.1^\circ$ for the 247 mergers at larger separations. The finite-separation sample contains 281 very minor, 37 minor, and 34 major mergers. The corresponding near/far counts are 76/205, 12/25, and 12/22, respectively. The 24 minor and major mergers in the near subset are distributed across 15 hosts, with no host contributing more than three. Both temporal subsets are numerically dominated by very minor mergers; the mass-stratified null tests in Sect.~\ref{app:merger_linked_boundaries} determine which classes show a timing excess relative to this abundance.

\subsection{Merger timing relative to episode boundaries and abrupt events}
\label{app:merger_linked_boundaries}

\begin{figure}[H]
\centering
\includegraphics[width=0.9\textwidth]{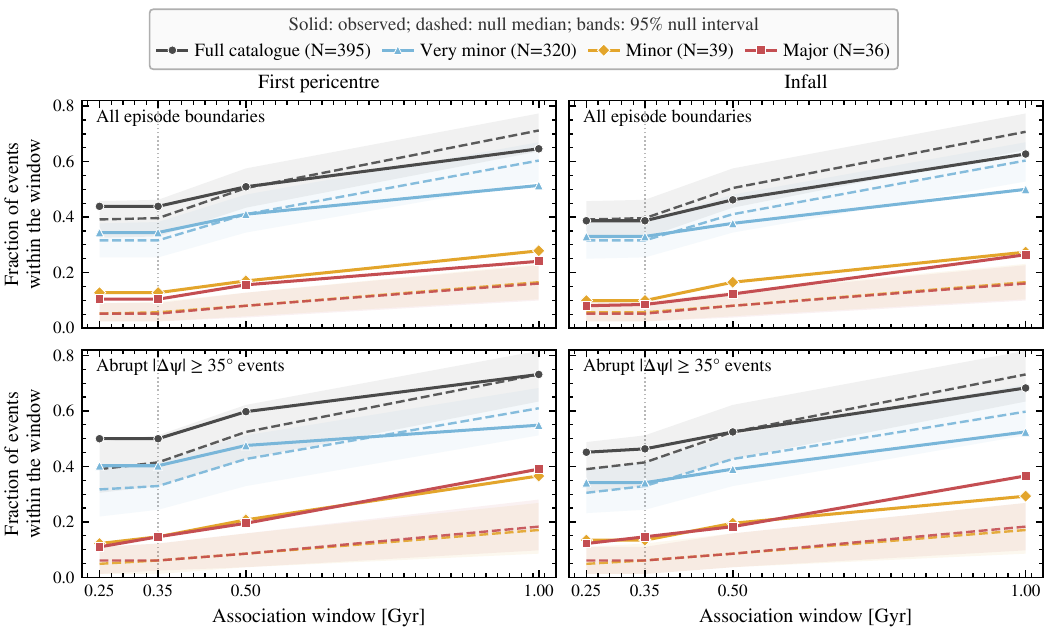}
\caption{Association of merger milestones with kinematic events. Columns show first pericentre and infall, while rows show all 212 episode boundaries and the 82 abrupt $|\Delta\psi|\geq35^\circ$ events. At each association window, the vertical axis gives the fraction of kinematic events lying within that interval of at least one milestone from the selected merger sample in the same galaxy. Solid curves give the observed fractions, dashed curves the medians of 5000 galaxy-preserving null realisations, and shaded bands their 95\% intervals. The null model randomises milestone times over the valid snapshot grid of each host while preserving its number of mergers in the selected class. Curves show the full catalogue ($N=395$), very minor ($N=320$), minor ($N=39$), and major ($N=36$) mergers; the mass intervals follow Sect.~\ref{subsec:merger_characterisation}. Vertical dotted lines mark the fiducial 0.35~Gyr window. Mass-class fractions are not additive because a kinematic event can be associated with more than one class.}
\label{fig:transition_null}
\end{figure}

Figure~\ref{fig:transition_null} compares the observed proximity of merger milestones to two sets of kinematic events: the 212 episode boundaries and the 82 abrupt $|\Delta\psi|\geq35^\circ$ events. For each association window, a kinematic event is counted when at least one infall or first-pericentre milestone from the selected merger class lies within that interval in the same galaxy. The observed fraction is the number of matched kinematic events divided by the total number in the corresponding row of the figure.

The null model redistributes the milestone times over the valid snapshot grid of each host while preserving the number of mergers of the selected mass class in every galaxy. We generate 5000 realisations and compare the observed fraction with the null median and its 95\% interval. The reported one-sided Monte Carlo probability is $(1+N_{\rm null,\geq obs})/5001$. The catalogue contains 320 very minor, 39 minor, and 36 major mergers, distributed across 42, 27, and 24 galaxies, respectively. The 75 minor and major mergers occur in 34 of the 44 galaxies.

At the fiducial 0.35~Gyr window, very minor mergers show no significant excess around abrupt events at either first pericentre ($p=0.074$) or infall ($p=0.458$). Minor mergers show an excess at both first pericentre ($p=2.0\times10^{-3}$) and infall ($p=9.2\times10^{-3}$), as do major mergers ($p=4.4\times10^{-3}$ and $p=4.0\times10^{-3}$, respectively). The timing excess around abrupt events is therefore confined to the minor and major mass classes; it is not driven by the numerically dominant very minor population.

For all episode boundaries, very minor mergers again show no excess at 0.35~Gyr, with $p=0.230$ at first pericentre and $p=0.370$ at infall. Minor mergers remain in excess at both milestones ($p=8.0\times10^{-4}$ and $p=1.7\times10^{-2}$), whereas major mergers show an excess at first pericentre ($p=8.4\times10^{-3}$) but not at infall ($p=0.063$). The full catalogue does not exceed its null expectation at the fiducial window because its chance-association rate is set by all 395 milestones and is dominated by very minor mergers. The full-catalogue fraction is not the sum of the three mass-class fractions, since the same kinematic event can lie within the selected window of mergers from more than one class.

The null model controls for the merger abundance and tracked time baseline of each galaxy. It does not preserve the cosmological clustering of merger and kinematic-event times; the measured excess therefore establishes temporal association, not direct causal coupling. The test pools transitions into and out of non-alignment.

\end{appendix}

\end{document}